\documentclass[aps,prb,twocolumn, superscriptaddress]{revtex4-2}
\usepackage{graphicx}
\usepackage{color}
\usepackage[usenames,dvipsnames,svgnames,table]{xcolor}
\usepackage{natbib}
\usepackage{epstopdf}
\usepackage{epsfig}
\usepackage{amsmath}
\usepackage{comment}
\usepackage{physics}
\usepackage{float}
\usepackage[caption=false]{subfig}
\newcommand{\beq}{\begin{equation}}
\newcommand{\eeq}{\end{equation}}
\newcommand{\bal}{\begin{aligned}}
\newcommand{\eal}{\end{aligned}}

\newcommand{\lp}{\left(}
\newcommand{\rp}{\right)}

\newcommand{\visc}[5]{#1^{#2\hphantom{#3}#4}_{\hphantom{#2}#3\hphantom{#4}#5}}
\newcommand{\bbnote}[1]{\textcolor{blue}{{\bf BB: #1}}}
\newcommand{\notepvr}[1]{\textcolor{red}{{\bf PVR: #1}}}
\begin{document}
\title{Resolving Hall and dissipative viscosity ambiguities via boundary effects}
\begin{abstract}

We examine the physical implications of the viscous redundancy of two-dimensional anisotropic fluids, where different components of the viscosity tensor lead to identical effects in the bulk of a system  [Rao and Bradlyn, Phys. Rev. X {\bf 10}, 021005 (2020)]. We first re-introduce the redundancy, show how it reflects a lack of knowledge of microscopic information of a system, and give microscopic examples. Next, we show that fluid flow in systems with a boundary can distinguish between otherwise redundant viscosity coefficients. 
In particular, we show how the dispersion and damping of gravity-dominated surface waves can be used to resolve the redundancies between both dissipative and Hall viscosities, and discuss how these results apply to recent experiments in chiral active fluids with nonvanishing Hall viscosity. Our results highlight the importance of divergenceless, magnetization-like contributions to the stress (which we dub ``contact terms'').  Finally, we apply our results to the hydrodynamics of quantum Hall fluids, and show that the extra contribution to the action that renders the bulk Wen-Zee action gauge invariant in systems with a boundary can be reinterpreted in terms of the bulk viscous redundancy.
\end{abstract}

\author{Pranav Rao}
\affiliation{Department of Physics and Institute for Condensed Matter Theory, University of Illinois at Urbana-Champaign, Urbana IL, 61801-3080, USA}

\author{Barry Bradlyn}%
\email{bbradlyn@illinois.edu}
\affiliation{Department of Physics and Institute for Condensed Matter Theory, University of Illinois at Urbana-Champaign, Urbana IL, 61801-3080, USA}

\maketitle
\section{Introduction}\label{sec:intro}
Viscosity in a fluid describes stresses developed in response to time-dependent strains. Viscous forces can be either dissipative or non-dissipative;
The former arise from the dissipative viscosity, which for an isotropic fluid consists of the familiar bulk and shear viscosities\cite{landau1987fluid}. 
The latter come from the time-reversal odd part of the viscosity tensor called the Hall (odd) viscosity\cite{1995-AvronSeilerZograf,1998-Avron}, which has been recently studied in topological phases\cite{haldane2009hall,hughesleighfradkin,bradlyn2012kubo,bradlyn2014low,gromov2017bimetric,gromov2016boundary,gromov2017investigating,robredo2021new,delacretaz2017transport,berdyugin2019measuring,link2018elastic,read2009non,read2011hall,Hoyos2012,shapourian2015viscoelastic,offertaler2019viscoelastic} and in classical chiral active fluids\cite{vitelli2017odd,soni2019odd,markovich2020odd,klymko2017statistical,shankar2020topological,souslov2020anisotropic,han2020statistical,souslov2019topological}.

In this work, we expand on and explore the experimental consequences of the \textit{viscous redundancy} highlighted in Ref.~\onlinecite{rao2019hall}: in anisotropic systems, there are more viscosity coefficients than independent bulk viscous forces. 
We show that this reflects a lack of experimentally accessible microscopic information in the bulk of the system. 
We derive several implications for our understanding of hydrodynamics in general, and resolve the redundancy by studying boundary phenomena. We show that the dispersion of boundary waves can be used as an experimental probe of viscosity coefficients that cannot be distinguished in the bulk. Furthermore, we will derive constraints on the degree to which power dissipation and angular momentum conservation can be used to extract individual viscosity coefficients, absent additional microscopic information. 

The viscosity determines fluid flow through the Navier-Stokes equation
\begin{equation}\label{eq:ns}
    \partial_t g_j + \partial_i(\tau^i_{\hphantom{i}j}+g_jv^i ) =0,
\end{equation}
where $g_j$ is the momentum density, $\mathbf{v}$ is the fluid velocity, and the stress tensor is
\begin{equation}
\tau^i_{\hphantom{i}j} = p\delta^i_j - \visc{\eta}{i}{j}{k}{l}\partial_kv^l,
\end{equation}
with pressure $p$. 
Roman indices index the Cartesian directions, and repeated indices are summed. For conceptual clarity we maintain the ``natural'' orientation of upper indices for velocity and lower indices for momentum. 
This is particularly important since the stress tensor $\tau^i_{\hphantom{i}j}$ is not symmetric for anisotropic fluids; the first index of the stress tensor denotes which surface on which internal forces act, while the second index denotes the direction of the force. We give a complete account of the notation used in this work in Appendix~\ref{app:notation}.

Since bulk flows are only sensitive to bulk viscous force 
\beq
f^\eta_{\text{bulk}, j} = \partial_i\visc{\eta}{i}{j}{k}{l}\partial_kv^l
\eeq 
rather than the individual viscosities themselves, this implies that the viscosity coefficients contain redundant information.

For the Hall viscosity, this can be seen in fluids with threefold or higher rotational symmetry, where the Hall viscosity tensor takes the form
\begin{align}\label{eq:hallviscgeneral_main}
\visc{(\eta^\mathrm{H})}{i}{j}{k}{l}=& \eta^\mathrm{H}\left(\delta^{ik}\epsilon_{jl}-\delta_{jl}\epsilon^{ik}\right) +\bar{\eta}^\mathrm{H}\left(\delta^i_j\epsilon^k_{\hphantom{k}l}-\delta^k_l\epsilon^i_{\hphantom{i}j}\right),
\end{align}
 where $\eta^\mathrm{H}$ is the isotropic Hall viscosity, $\bar{\eta}^\mathrm{H}$ is a second angular-momentum nonconserving Hall viscosity, $\delta$ is the Kronecker delta, and $\epsilon$ is the antisymmetric Levi-Civita symbol.
This leads to a bulk viscous force
\beq
{\bf f}^{\mathrm{H}}_\mathrm{bulk} =  (\eta^\mathrm{H}+\bar{\eta}^\mathrm{H})\nabla^2{\bf v}^*,
\eeq
 where we have defined 
 \beq
 v^{*,i} = \epsilon^{i}_j v^j.
 \eeq 
The bulk viscous force is determined by the sum 
\beq
\eta^{\mathrm{H}}_\mathrm{tot} = \eta^H + \bar{\eta}^H
\eeq
of the two Hall viscosity coefficients, and so $\eta^\mathrm{H}$ and $\bar{\eta}^\mathrm{H}$\cite{rao2019hall,souslov2019anisotropic} are \textit{redundant}, as they have the same effect in the bulk equations of motion. 
The difference 
\beq
\eta^\mathrm{H}_\mathrm{diff} = \eta^\mathrm{H}-\bar{\eta}^\mathrm{H}
\eeq
does not enter into the bulk force, and can be shifted by a ``contact''\cite{rao2019hall} term 
\beq
\delta \tau^i_{\hphantom{i}j} = C_0\partial^{*i} v_j
\eeq 
in the bulk stress tensor, as shown explicitly in Appendix~\ref{app:viscreview}. 
As $\delta\tau^i_{\hphantom{i}j}$ can be written as the curl of a vector, we can view the contact term  as a ``magnetization stress"~\cite{cooper1997thermoelectric,bradlyn2014low}. 
Analogous to electrical magnetization, a uniform magnetization stress has no effect in the bulk but will give rise to a force on the boundary; thus we expect the boundary force to depend on $\eta^\mathrm{H}_\mathrm{diff}$. This magnetization stress has previously been interpreted as a ``torsional Hall viscosity''~\cite{hughesleighfradkin,hoyos2014hall,bradlyn2014low}, which we revisit in Appendix~\ref{app:ambiguities}.

 Similar considerations also apply to the dissipative viscosity. 
 For an incompressible anisotropic fluid with threefold (or higher-than-fourfold) rotational symmetry, the dissipative viscosity is\cite{rao2019hall,cook2019electron,cook2021viscometry}
\begin{align} \label{eq:dissipativevisc}
   \visc{(\eta^\mathrm{D})}{i}{j}{k}{l} =\eta^\mathrm{sh}\visc{(\sigma^x\odot\sigma^x+\sigma^z\odot\sigma^z)}{i}{j}{k}{l}+\eta^\mathrm{R}\visc{(\epsilon\odot\epsilon)}{i}{j}{k}{l}, \nonumber 
    \end{align}
where $\odot$ is the symmetric tensor product. Note that, as discussed in Appendix~\ref{app:viscreview}, in general there can also be an additional contribution $\eta^{\mathrm{RC}} \propto \delta\odot\epsilon$ to the viscosity tensor. 
However, for an incompressible fluid this cannot be distinguished from $\bar{\eta}^\mathrm{H}$, and so we will set $\eta^{\mathrm{RC}}=0$ for convenience. 
See Ref.~\cite{monteiro2021hamiltonian} for a related discussion. 
Furthermore, we exclude the unique case of fourfold rotational symmetry where the shear viscosity splits.
The bulk dissipative viscous force is proportional to the sum of dissipative viscosities, 
\beq
{\bf f}^\mathrm{dis}_\mathrm{bulk} = \left(\eta^\mathrm{sh} + \eta^\mathrm{R}\right) \nabla^2 {\bf v},
\eeq
illustrating that $\eta^\mathrm{sh}$ and $\eta^\mathrm{R}$ are redundant. 
The dissipative contact term (derived in detail in Appendix~\ref{app:viscreview_diss}) 
\beq
\delta \tau^i_{\hphantom{i}j} = C_{\mathrm{dis}}(\partial_j v^i-\delta^i_j\nabla\cdot\mathbf{v}) 
\eeq
gives a magnetization stress that
shifts the difference 
\beq
\eta^{\mathrm{dis}}_{\mathrm{diff}} =\eta^{\mathrm{sh}}-\eta^\mathrm{R}
\eeq
 while leaving the sum
 \beq \label{eq:distotdef}
\eta^\mathrm{dis}_\mathrm{tot}  = \eta^\mathrm{sh} + \eta^\mathrm{R}
\eeq
 fixed. 
Consequently, considering dissipation $\mathbf{f}^\mathrm{dis}\cdot\mathbf{v}$ and using only the bulk equations of motion\cite{landau1987fluid} requires $\eta^{\mathrm{dis}}_{\mathrm{tot}}>0$; microscopic information other than the flow is necessary to say more\cite{cook2021viscometry}. In Appendix~\ref{app:power} we derive this explicitly by computing the dissipated power from the mass and momentum continuity equations.

Unless the bulk stress tensor is directly measurable (which requires knowledge of microscopics), redundant viscosity coefficients are indistinguishable through bulk flow measurements, which probe the force $f_{\mathrm{bulk},j}^\eta$. 
As an example, we note that if the internal angular momentum $L_\mathrm{int}$ and its associated flux $M^{\mathrm{int},k}$ are known (e.g. in a liquid crystal\cite{parodi} or spin-orbit coupled electron fluid\cite{denisov2022spin}), then for an incompressible fluid angular momentum conservation
\begin{equation}
\label{eq:angmomconst_main}
    \partial_t L_\mathrm{int}({\bf r},t) = \epsilon^j_i\tau^i_{\hphantom{i}j} + \partial_k M^{\mathrm{int},k}
\end{equation}
determines the antisymmetric part of the stress tensor. 
Combined with Eq.~\eqref{eq:ns}, this determines  $\eta^\mathrm{H}-\bar{\eta}^\mathrm{H}$ and $\eta^{\mathrm{sh}}-\eta^\mathrm{R}$, as  shown in Appendix~\ref{app:ambiguities}. 
For example, in a fluid with no internal angular momentum,  Eq.~\eqref{eq:angmomconst_main} requires the antisymmetric stress to vanish, telling us that $\eta^\mathrm{R}=\bar{\eta}^\mathrm{H}=0$. 
For a non-rotationally-invariant example, in Appendix~\ref{eq:app_massaniso} we contrast a quantum Hall fluid with mass anisotropy and an isotropic quasi-2D quantum Hall fluid in a tilted magnetic field. The low-energy spectrum for these two systems are identical~\cite{yang2017anisotropic}. Furthermore these two systems have the same bulk Hall viscous forces. However, as we review in Appendix~\ref{eq:app_massaniso}, the two systems have different viscosity tensors, due to the magnetization stresses~\cite{offertaler2019viscoelastic,gromov2017investigating}. Absent knowledge of the microscopic Hamiltonian, bulk flow measurements  cannot distinguish between these two systems. 
Thus, the viscous redundancy can be experimentally resolved, and the value of the magnetization stress fixed, if the microscopic internal degrees of freedom of the fluid can be directly measured or inferred.

Similarly, Ref.~\cite{cook2021viscometry} argued that the redundancy in the dissipative viscosity can be resolved through measurements of local heating. 
As we show in Appendix~\ref{app:power}, however, this  relies on knowing the microscopic form of the bulk energy current. 
Without this microscopic knowledge, the local heating rate suffers from the same ambiguity as the force density, essentially since the local heating rate in the bulk is proportional to $\mathbf{f}^\eta_{\mathrm{bulk}}\cdot\mathbf{v}$. 
For the same reason, the local heating rate cannot be used as a probe of the Hall viscosity.

In the absence of experimental or theoretical access to microscopic information, we can use boundary effects to resolve the viscous ambiguity, as redundant viscosity coefficients provide unique forces on a fluid boundary. 
Independently, viscous boundary effects have been an interesting area of study\cite{lamb1924hydrodynamics}, especially for the Hall viscosity\cite{soni2019odd,abanov2018odd,abanov2019free,abanov2020hydrodynamics,ganeshan2017odd,Wiegmann_boundary,Wiegmann_vortex}. 
The Hall viscosity $\eta^\mathrm{H}$ is often viewed as ``trivial" in the bulk of an incompressible fluid, since it can be absorbed into a redefinition of the pressure; on the boundary it provides a nontrivial effect\cite{Wiegmann_boundary,Wiegmann_vortex}. 
In field theories of hydrodynamics\cite{abanov2019free}, Hall viscosity is encoded in geometric terms in the bulk and boundary action for the fluid\cite{gromov2016boundary}. 
We will show how this reflects the redundancy between $\eta^\mathrm{H}$ and $\bar{\eta}^\mathrm{H}$. 
From an experimental perspective, boundary effects of $\eta^\mathrm{H}$ have been studied through free surface waves\cite{abanov2018odd,abanov2020hydrodynamics}, culminating in one of the first measurements of the Hall viscosity in a colloidal chiral fluid\cite{soni2019odd}.

Here we consider boundary effects to resolve the viscous ambiguity. We interpret magnetization stresses through the language of stress boundary conditions, and describe a trade-off between modified boundary conditions and no-stress boundary conditions with a bulk contact term. 
We show how surface wave dispersion relations disambiguate redundant viscosities, both dissipative and non-dissipative. 
We relate this to recent experiments examining Hall viscosity in chiral active fluids. 
We then revisit the effective description of quantum Hall fluids, showing how the boundary term added to the action to preserve gauge invariance\cite{gromov2016boundary} can be interpreted as a (gauge-noninvariant) magnetization stress, revealing a new perspective on this system.

\section{Boundary Forces}

We begin by computing the boundary viscous forces for an incompressible fluid. 
The boundary Hall viscous force is 
\beq
f^\mathrm{H}_{\mathrm{bdd},j}= \hat{n}_i (\tau^\mathrm{H})^i_{\hphantom{i}j},
\eeq 
with $n_i$ the unit boundary normal vector. 
For an incompressible fluid, we find from Eq.~\eqref{eq:hallviscgeneral_main} that
\beq
\bal
\label{eq:nondissbf}
{\bf f}^\mathrm{H}_\mathrm{bdd}&=  \left[\left(\eta^\mathrm{H}_\mathrm{tot} + \eta^\mathrm{H}_\mathrm{diff}\right) \left( \partial_{\bf s} v_{\bf n} + \frac{v_{\bf s}}{R} \right) + \eta^\mathrm{H}_\mathrm{tot} \omega \right] {\bf \hat{n}} \\
&+ \left[\left(\eta^\mathrm{H}_\mathrm{tot} + \eta^\mathrm{H}_\mathrm{diff}\right) \left( \partial_{\bf s} v_{\bf s} - \frac{v_{\bf n}}{R} \right)\right]{\bf \hat{s}},
\eal
\eeq
where $\hat{\bf s} = -\hat{\bf n}^*$ is the boundary tangent vector, $\omega = \epsilon^i_{\hphantom{i}j} \partial_i v^j$ is the vorticity, and $R = 1/\kappa$ is the local radius of curvature of the boundary\cite{gromov2016boundary,soni2019odd}. 
The pressure-like contribution $\eta^\mathrm{H}_\mathrm{tot}\omega\hat{\bf n}$ in Eq.~\eqref{eq:nondissbf} is the bulk force restricted to the boundary and can be captured by defining the modified pressure\cite{ganeshan2017odd, abanov2018odd,abanov2019free} $
\tilde{p} = p - \eta^\mathrm{H}_\mathrm{tot}\omega$. 
This reflects a more general sentiment from previous works that the only bulk effect of the Hall viscosity is to modify the pressure\cite{Wiegmann_boundary,Wiegmann_vortex,soni2019odd}. 
We see from Eq.~(\ref{eq:nondissbf}) that the boundary force has additional terms, including contributions dependent on $\eta^\mathrm{H}_\mathrm{diff}$ and therefore on $C_0$.

Analogously, the boundary dissipative viscous force is
\beq
\bal
\label{eq:dissbf_main}
{\bf f}^\mathrm{dis}_\mathrm{bdd}&=  \left[\left(\eta^\mathrm{dis}_\mathrm{tot} + \eta^\mathrm{dis}_\mathrm{diff}\right) \partial_{\bf n} v_{\bf{n}} \right] {\bf \hat{n}} \\
&+ \left[\eta^\mathrm{dis}_\mathrm{tot} \omega + (\eta^\mathrm{dis}_\mathrm{tot}+\eta^\mathrm{dis}_\mathrm{diff}) \left(\partial_{\bf n} v_{\bf s} - \frac{v_{\bf s}}{R}\right) \right]{\bf \hat{s}}.
\eal
\eeq
The boundary force depends on both the bulk observable $\eta^\mathrm{dis}_\mathrm{tot}$, and the difference $\eta^\mathrm{dis}_\mathrm{diff}$.
In order for the differences $\eta^\mathrm{H}_\mathrm{diff} \; \text{and} \; \eta^\mathrm{dis}_\mathrm{diff}$ to have a measurable effect on flows, we must consider systems with a boundary.

\subsection{Stress Boundary Conditions} 

We now relate the viscous redundancy to  boundary conditions on the stress tensor\cite{lamb1924hydrodynamics,ganeshan2017odd,kiselev2019boundary,delacretaz2017transport}. 
The no-stress boundary condition, relevant for the free surface considered later, is given by, 
\beq\label{eq:nostressbcs}
\hat{n}_i \tau^{i}_{\hphantom{i}j} = -p{\hat{n}}_j+{f}^\mathrm{H}_{\mathrm{bdd},j} + {f}^\mathrm{dis}_{\mathrm{bdd,j}} =0,
\eeq
for a fluid with pressure $p$. For the sake of brevity we ignore surface tension, which would modify the tangential component of Eq.~\eqref{eq:nostressbcs} 
We see that both the tangent and normal components of Eq.~\eqref{eq:nostressbcs} depend on the differences $\eta^\mathrm{dis}_\mathrm{diff}$ and $\eta^\mathrm{H}_\mathrm{diff}$, and therefore the no-stress boundary conditions are sensitive to the magnetization stresses and in turn to the contact terms $C_0$ and $C_\mathrm{dis}$.

The normal component of Eq.~\eqref{eq:nostressbcs} requires that the modified pressure balance the viscous forces at the boundary. It was previously thought that balancing the tangential no-stress condition for an isotropic fluid without dissipation would require finite curvature $R$\cite{abanov2018odd,soni2019odd}. 
Here in contrast, we see the tangential component in Eq.~\eqref{eq:nostressbcs} can be balanced if $\eta^\mathrm{H}_{\mathrm{diff}} = -\eta^\mathrm{H}_{\mathrm{tot}}$ even if $R\rightarrow \infty$.

The boundary force allows us to probe the values of $\eta^\mathrm{H}_\mathrm{diff}$ and $\eta^\mathrm{dis}_\mathrm{diff}$, which we can view as intrinsic properties of the fluid. 
Alternatively, changes to the difference viscosities (and hence changes to the magnetization stress) can be absorbed into a modification of the boundary conditions. 
This trade-off reflects a broader statement that changing the microscopic definition of the magnetization stress in the bulk modifies the notion of no-stress boundary conditions.  
This has implications for unambiguously determining the viscosity coefficients when the boundary conditions are not well controlled (as in electron hydrodynamics) and when the bulk stress tensor cannot be directly measured. 
We will see a concrete example of this trade-off in quantum Hall fluids. 
First we explore the implications of Eq.~\eqref{eq:nostressbcs} for surface waves.

\section{Surface Waves}\label{sec:waves}

The viscous redundancy can be translated into a physical effect by considering surface waves on an incompressible, anisotropic fluid with Hall viscosity. 
We consider linearized waves on the surface of a half plane with height $h(x,t)$, in the presence of a gravitational field $-g{\bf \hat{y}}$. 
A linearized kinematic boundary condition 
\beq
\partial_t h = v_y(y=h),
\eeq
ensures the continuity of the velocity at the boundary.
We also have the no stress condition Eq.~\eqref{eq:nostressbcs}, where to linear order 
\begin{align}
\mathbf{\hat{n}} &= \mathbf{\hat{y}}, \\
\mathbf{\hat{s}} &= -\mathbf{\hat{x}}. 
\end{align}
We take a wave ansatz 
\beq
\mathbf{v}\propto\exp[i(kx-\Xi t)]
\eeq 
and solve for the dispersion 
\beq
\Xi(k) = \xi(k) - i\Gamma(k),
\eeq
where $\xi(k)$ is the frequency and $\Gamma(k)$ the damping rate.

\subsection{Gravity-dominated waves}

When the gravitational force is the dominant scale, we can follow the approach of Ref.~\cite{abanov2018odd}. We introduce the dimensionless parameter 
\beq
\beta^2 = {\eta^\mathrm{dis}_\mathrm{tot} k^2}/{\sqrt{gk}}.
\eeq
Gravity dominated waves occur when $\beta << 1$, 
with all viscosities treated as small comparatively.
In Appendix~\ref{sec:two} we show that in this limit the dispersion relation for surface waves is given by
\beq\label{eq:gwavexi}
\bal
\xi_\pm(k) = \pm \sqrt{gk} - 2\eta^\mathrm{H} k^2,\;\;
\Gamma_\pm(k) = 2\eta^\mathrm{sh} k^2.
\eal
\eeq
We see that $\Gamma_\pm(k)$ depends on $\eta^\mathrm{dis}_\mathrm{diff}$ through $\eta^\mathrm{sh}$, whereas $\xi_\pm(k)$ depends on $\eta^\mathrm{H}_\mathrm{diff}$ via $\eta^\mathrm{H}$. 
Note that we obtain the same results to leading order in $k$, even without considering the viscosities to be small. 
Eq.~\eqref{eq:gwavexi} agrees with Ref.~\cite{abanov2018odd} despite our additional nonzero viscosity coefficients.

We thus propose that the damping rate gives an experimental measure of the difference between dissipative viscosities,
\begin{equation}
\label{eq:surfacewavedissipation}
    \frac{\Gamma_\pm}{k^2} - \eta^\mathrm{dis}_\mathrm{tot} = \eta^\mathrm{dis}_\mathrm{diff}.
\end{equation}
Similarly, $\xi$ can be used to experimentally measure the difference between non-dissipative viscosities,
\beq\label{eq:surfacewavedisp}
\eta^\mathrm{H}_\mathrm{diff} = \pm \frac{\sqrt{g}}{k^{3/2}} - \frac{\xi_\pm}{k^2} - \eta^\mathrm{H}_\mathrm{tot}.
\eeq
Recall that $\eta^\mathrm{H}_\mathrm{tot}$ and $\eta^\mathrm{dis}_\mathrm{tot}$ can in principle be determined from independent \emph{bulk} measurements: Eqs.~\eqref{eq:surfacewavedissipation} and \eqref{eq:surfacewavedisp} allow us to determine $\eta^\mathrm{H}_\mathrm{diff}$ and $\eta^\mathrm{dis}_\mathrm{diff}$, and therefore resolve the viscous ambiguity. 

It is possible for the damping $\Gamma_\pm=0$ even in a dissipative fluid provided $\eta^\mathrm{dis}_\mathrm{diff} \rightarrow -\eta^\mathrm{dis}_\mathrm{tot}$, i.~e. if all dissipation is due to rotational viscosity. 
Alternatively, we can get the same result for a fluid with no rotational viscosity by viewing the magnetization stress as modifying the no stress boundary conditions; it is only when the boundary conditions are fixed that $\Gamma$ resolves the dissipative ambiguity. 
This can also be viewed as a modification of Eq.~\eqref{eq:nostressbcs}, interpreting $C_\mathrm{dis}$ as an anomalous stress at the boundary.

Furthermore, when $\eta^{sh} < 0$ our surface waves grow exponentially in time. 
This implies that the fluid surface is unstable at the linearized level. 
Thus non-negativity of the shear viscosity alone is dictated by stability of the free surface, while the bulk equations of motion require $\eta^\mathrm{dis}_\mathrm{tot} \geq 0$; there is no further constraint on  $\eta^R$ from this setup.

\subsection{Chiral Viscosity Waves}

We next consider $g=0$ and find chiral waves propagating along the boundary of the half plane, in agreement with previous work\cite{soni2019odd, ganeshan2017odd}. The details of the calculation are given in Appendix~\ref{sec:two_chiral}.  
To leading order in $\eta^\mathrm{dis}_{\mathrm{tot}}$, the dispersion is given by
\beq 
\Xi = -2 \eta^\mathrm{H} k^2  - 2i k^2 \sqrt{|\eta^\mathrm{H}| \eta^\mathrm{dis}_\mathrm{tot}}.
\eeq
This indicates that the chiral waves move in a direction set by the Hall viscosity. 
Importantly, it is only the $\textit{component}$ $\eta^\mathrm{H}$ rather than $\eta^\mathrm{H}_\mathrm{tot}$ that sets the direction. 
Hence the direction of the waves cannot be determined from bulk data alone, or equivalently that the expression above is sensitive to the non-dissipative contact term. Because we assumed that the dissipative viscosities were small in this derivation, we do not find a dependence of the dispersion relation on the dissipative contact term at this order.

\subsection{Chiral Active Fluids}\label{sec:wave_fluid} 
\begin{figure}
    \centering
    \includegraphics[scale=.25]{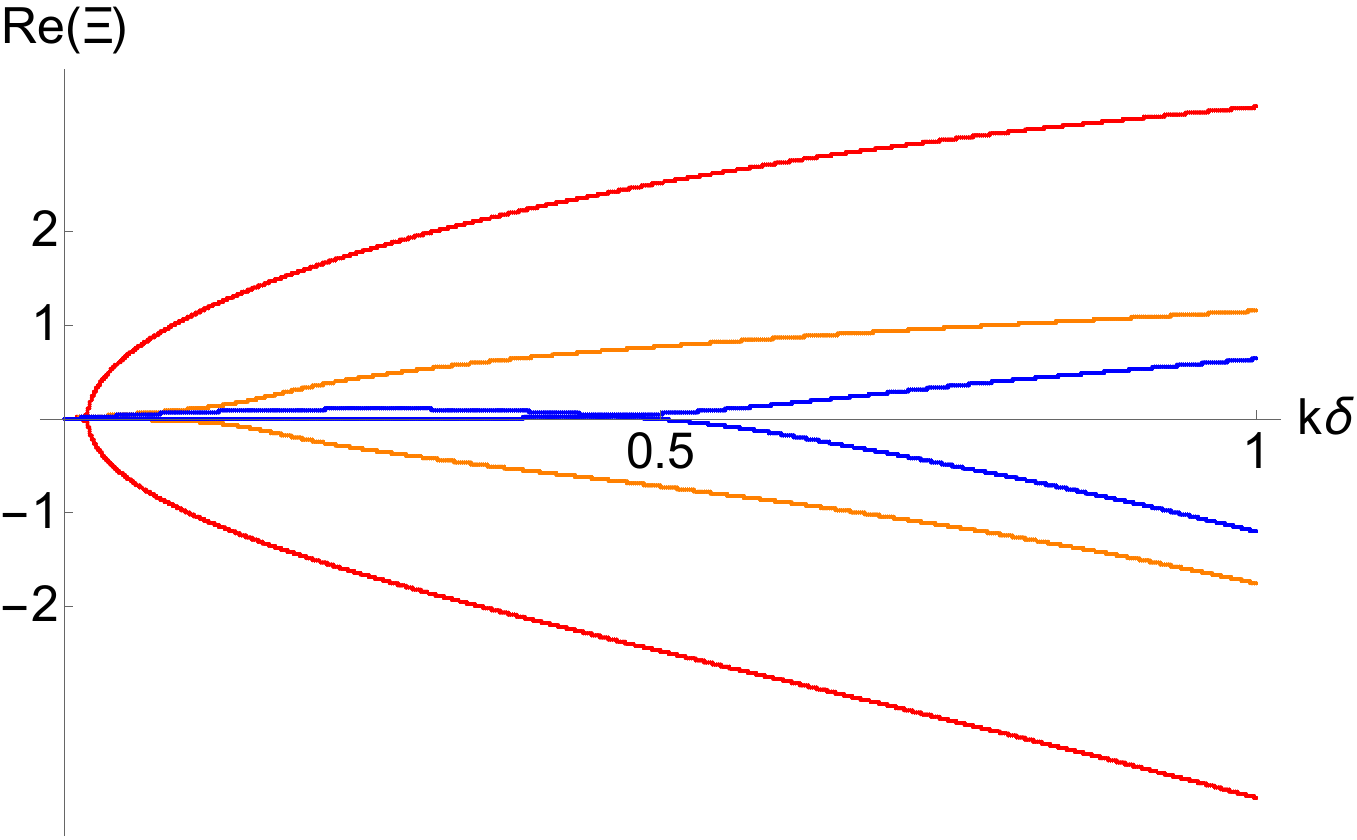}
    \caption{Dispersion relation for surface waves with gravity and time-reversal breaking from a local rotation rate $\Omega$. 
    The red plot has $g=10$, the blue plot has $g=1$ and the orange has $g=1.2$. 
    The other parameters are fixed at $\eta^\mathrm{sh}=0.1,\eta^\mathrm{R}=0.5,\eta^\mathrm{H}=0.3, \Omega=-0.6$ and $\mu=1$. 
    We see that as $g$ increases, the dispersion relations begin to converge to the Lamb wave dispersion of $\pm \sqrt{gk}$.}
    \label{fig:dispgneq0}
\end{figure}

So far we have considered a fluid with an external mechanism of time-reversal symmetry breaking, such as a magnetic field. 
Recent experiments on colloidal chiral active fluids, however, break time-reversal via a local rotation rate $\Omega$ for fluid particles\cite{soni2019odd}. 
This changes the constitutive relation for the stress tensor to measure vorticity as a deviation from $2\Omega$. 
As shown in Appendix~\ref{app:active}, this allows for a steady state vorticity which takes the value $\omega_s = \eta^\mathrm{R} \Omega/\eta^\mathrm{dis}_\mathrm{tot}$ at $y=0$. 
We also introduce a frictional force $\mu$ between the fluid and the substrate, which introduces a hydrodynamic length $\delta = \sqrt{\eta^\mathrm{dis}_\mathrm{tot}/\mu}$. 
In the long wavelength $k\delta << 1$ limit where gravity is small compared to other scales, we show in Appendix~\ref{app:active} that there are two physical modes 
\beq
\bal
\Xi_{1g}(k) &=2(i\eta^\mathrm{H} - \eta^\mathrm{sh}) \frac{2\Omega \delta \eta^\mathrm{R}}{i \eta^\mathrm{dis}_\mathrm{tot}} k^3 - \frac{i g k \delta}{\sqrt{\eta^\mathrm{dis}_\mathrm{tot} \mu}}  \\
\Xi_{2 g}(k) &= -i\mu -\frac{2\Omega\eta^R}{\eta_\mathrm{tot}^\mathrm{dis}}k\delta + \frac{i g k \delta}{\sqrt{\eta^\mathrm{dis}_\mathrm{tot} \mu}}.
\eal
\eeq
The $\Xi_{2,g}$ mode is strongly overdamped at small $k$. 
Despite the inclusion of the additional Hall viscosity $\bar{\eta}^\mathrm{H}$, the $\Xi_{1,g=0}(k)$ mode matches the dispersion relation found in Ref.~\cite{soni2019odd} in the absence of gravity. 
We see that the fluid surface is stable only if $\text{sign}(\eta^\mathrm{H} \eta^\mathrm{R} \Omega)<0$, in order to ensure perturbations decay exponentially in time. 
We see that the $\Xi_{1,g=0}$ mode is sensitive to contact terms via
\beq
\bal
\xi_{1,g=0}(k) &= -\left((\eta^\mathrm{dis}_\mathrm{tot})^2 -(\eta^\mathrm{dis}_\mathrm{diff})^2\right) \frac{\Omega \delta k^3}{\mu \eta^\mathrm{dis}_\mathrm{tot}}\\
\Gamma_{1,g=0}(k) &= -(\eta^\mathrm{H}_\mathrm{tot}+\eta^\mathrm{H}_\mathrm{diff})(\eta^\mathrm{dis}_\mathrm{tot}-\eta^\mathrm{dis}_\mathrm{diff})\frac{\Omega \delta k^3}{\mu \eta^\mathrm{dis}_\mathrm{tot}}.
\eal
\eeq
Finally, we note that there is a crossover to gravity-dominated waves for sufficiently large $g$ ($\beta\ll 1$):
\beq
\Xi_{1/2,g}\rightarrow \xi_{\mp}-i\Gamma_{\mp} -(i\mu+k\delta\omega_s)/2,
\eeq
 with $\xi_{\pm},\Gamma_{\pm}$ from Eq.~\eqref{eq:gwavexi}. 
 We show the dispersion for various $g$ in Fig.~\ref{fig:dispgneq0}.

\section{Quantum Hall Regime} \label{sec:qhregime}

Finally, we examine quantum Hall fluids. 
The quantum Hall fluid is dissipationless and rotationally invariant, and thus we only consider the isotropic Hall viscosity. Although the quantum Hall fluid is rotationally invariant, we will see that there are boundary forces which have the appearance of magnetization stress. In this section we work in 2+1 dimensions and use Greek works for spacetime indices (see Appendix~\ref{app:notation}). The Hall viscosity is given by\cite{1995-AvronSeilerZograf,Levay1995,read2009non} 
\beq
\eta^\mathrm{H}_{\mathrm{WZ}} = \frac{\nu \bar{s}}{4\pi} B,
\eeq
where, $\nu$ is the filling fraction\footnote{or equivalently the level of the Chern-Simons theory describing the quantum Hall phase}, $-\bar{s}$ is the average orbital spin per particle and $B$ is the magnetic field. 
The Hall viscosity derives from the Wen-Zee (WZ) action\cite{wen1992shift}
\beq
\label{eq:wz1}
S_\text{WZ} = \frac{\nu \bar{s}}{2\pi} \int_\mathcal{M} A \wedge d\bar{\omega}.
\eeq
The WZ term couples geometry ($SO(2)$ spin connection $\bar{\omega}$) to the $U(1)$ electromagnetic vector potential $A$. 
Absent a boundary, the variation of $S_{\text{WZ}}$ with respect to the geometry with fixed (reduced) torsion\cite{bradlyn2014low,Gromov20141} yields the bulk Hall viscous stress. 
To see this, we consider a strain perturbation $e^a_\mu = \delta^a_\mu + u^a_\mu(t)$ with traceless spatially-uniform deformation tensor $u^a_\mu = \partial_
\mu u^a(t)$. 
The nonvanishing component of the spin connection is\cite{bradlyn2014low,Gromov20141,hoyos2014hall}
$\bar{\omega}_0 = \frac{1}{2}\epsilon^{ab}e^\mu_a\partial_t e_\mu^b,$
and the corresponding bulk stress response is
\beq
\label{eq:isoviscstresstensor}
(\tau^{\text{WZ}})^i_{\hphantom{i}j} =\eta^\mathrm{H}_\mathrm{WZ}\left(\partial^i v^*_j + \partial^{*\, i} v_j \right).
\eeq
With a boundary present, the Wen-Zee action Eq.~\eqref{eq:wz1} is no longer invariant under $U(1)$ gauge transformations of the vector potential, and to preserve gauge invariance we must add the boundary action\cite{gromov2016boundary} 
\beq
\label{eq:boundaryterm}
S_\text{BT} = \frac{\nu \bar{s}}{2\pi} \int_{\partial M} A \wedge K,
\eeq
where the extrinsic curvature one-form $K= n_\mu \partial_\lambda s^\mu dx^\lambda$~\cite{gromov2016boundary,abanov2019free}. 
Eqs.~\eqref{eq:wz1} and \eqref{eq:boundaryterm} combine to yield the fully gauge invariant action 
\beq
\bal
\label{eq:qhactionfinal}
S &= \frac{\nu \bar{s}}{2\pi} \int_\mathcal{M} \bar{\omega} \wedge dA
-  \frac{\nu \bar{s}}{2\pi} \int_{\partial M} A \wedge d\alpha.
\eal
\eeq
Above, $\alpha$ is the angle between the boundary frame $\lbrace \bf n, s \rbrace$ and $\left.e_\mu^a\right|_{\partial\mathcal{M}}$ \footnote{the spin connection projected to the boundary satisfies the relation $\omega_\mu + K_\mu = \partial_\mu  \alpha$ (See Ref.~\onlinecite{gromov2016boundary} for more details)}. 

The first term in Eq.~\eqref{eq:qhactionfinal} is equivalent to Eq.~\eqref{eq:wz1} in the bulk. 
The bulk stress response is therefore given by Eq.~\eqref{eq:isoviscstresstensor}. 
The first term in Eq.~\eqref{eq:qhactionfinal} does not contribute to the boundary stress tensor. 
However, for a half plane geometry Ref.~\onlinecite{abanov2019free} showed that the second term in Eq.~\eqref{eq:qhactionfinal} gives a viscous force 
\beq
f^{\text{BT}}_n = -2\eta^\mathrm{H}_{\mathrm{WZ}} \partial_{\bf s} v_n
\eeq
normal to the boundary that modifies the boundary conditions.
\begin{figure}[t]
\label{fig:qh}
\begin{center}
\includegraphics[width=\columnwidth]{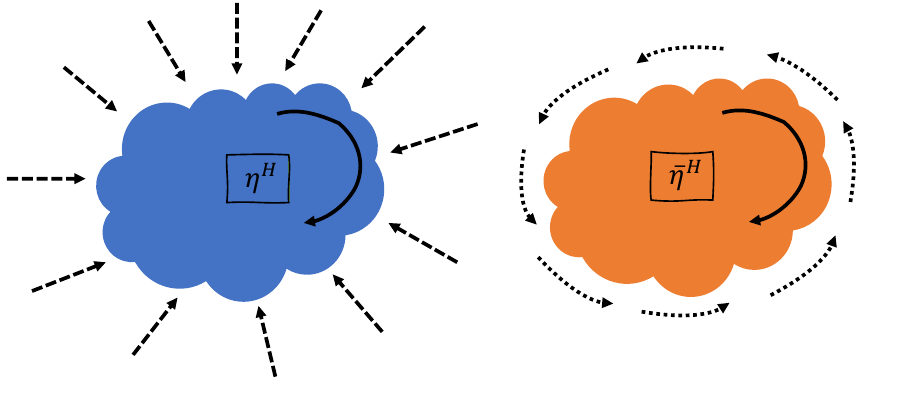}
\caption{Schematic of the two views of quantum Hall fluids presented. 
Left: fluid with Hall viscosity and a modified normal stress at the boundary and Right:  fluid with Hall viscosity and a bulk contact term, with zero normal stress at the boundary.} 
\end{center}
\end{figure}
We have chosen the gauge $A = -By dx$~\cite{abanov2018odd}. 
The total boundary force is now $\hat{n}^i \tau^\mathrm{WZ}_{ij}+f_n^{\mathrm{BT}}\hat{n}_j$. 
We may interpret the boundary term in Eq.~\eqref{eq:qhactionfinal}---and hence the boundary force---as arising from a contact term, choosing (in this gauge) $C_0 = -2\eta_\mathrm{WZ}^\mathrm{H}$. 
To this end, we can reinterpret the stress tensor of the system with the contact term added as
\beq
\label{eq:effectivestress}
(\tau^{WZ})^{i}_{\hphantom{i}j} + (\tau^{C_0})^i_{\hphantom{i}j}= \eta^\mathrm{H}_\mathrm{WZ} \left(\partial^i v^*_j - \partial^{*\, i} v_j \right).
\eeq
The stress tensor is no longer symmetric, and appears to break $U(1)$ gauge invariance in the bulk. 
The effective stress Eq.~\eqref{eq:effectivestress} reproduces the normal boundary force $\hat{n}^i\hat{n}^j \tau^\mathrm{WZ}_{ij}+f_n^{\mathrm{BT}}$ with a modification to the (already non-universal) tangential boundary condition. 
This is depicted in Fig.~2. 
The stress tensor Eq.~\eqref{eq:effectivestress}, corresponds to viscosities $\eta^\mathrm{H}=0, \bar{\eta}^\mathrm{H} = \eta^\mathrm{H}_\mathrm{WZ}$: all of $\eta^\mathrm{H}_\mathrm{tot}$ comes from the rotational symmetry-breaking coefficient $\bar{\eta}^H$. 
In the language of Refs.~\cite{abanov2019free,monteiro2021hamiltonian}, this means the boundary term has the effect of shifting the Hall viscosity into the ``odd pressure'' $\bar{\eta}^\mathrm{H}$. 
Rotational symmetry is restored by the additional tangential boundary force $-\hat{n}^i\hat{s}^j\tau^{C_0}_{ij}$. 

\section{Outlook}
We have seen that waves at a free surface provide an experimentally accessible way to distinguish between (dissipative and Hall) viscosity coefficients that produce identical bulk flows. Our work is directly applicable to experiments in chiral active fluids. Additionally, as showed in detail in Appendix~\ref{app:ambiguities}, the nondissipative magnetization stress is intimately related to ``torsional Hall viscosity''\cite{hughesleighfradkin,hoyos2014hall}. Our results thus serve as a guide to probing torsional response in two-dimensional fluids. 

Going forward, our approach extends to fluids with twofold rotational symmetry, where additional anisotropic viscosities appear. Additionally, exploring surface waves in compressible fluids could be relevant for both classical active fluids and composite Fermi liquid states. For compressible fluids, the dissipative viscous redundancy involves $\eta^\mathrm{sh},\eta^\mathrm{R},$ and the bulk viscosity $\zeta$, as we discuss in Sec.~\ref{app:viscreview_diss}. We expect that the interplay between Hall viscosity and odd torque\cite{monteiro2021hamiltonian} will play a larger role in the free surface properties of compressible fluids. Our analysis can be straightforwardly generalized to analyze partial-slip boundary conditions\cite{delacretaz2017transport} relevant for electron hydrodynamics and quantum Hall transport. Lastly, our work highlights the importance of local spin imaging\cite{denisov2022spin} as a probe of the viscous redundancy in electron fluids.

\begin{acknowledgments}
The authors thank S.~Ganeshan and A.~Abanov for helpful discussions. 
We also thank A.~Lucas for bringing Ref.~\cite{cook2021viscometry}, which presents an alternative method to disambiguate between redundant dissipative viscosities. 
This work was supported by the Alfred P. 
Sloan Foundation, and the National Science foundation under grant DMR-1945058. 
\end{acknowledgments}

\appendix
\onecolumngrid
\section{Notation and Conventions}\label{app:notation}

We denote two dimensional spatial indices by letters $i,j,k,l,...$ which label the Cartesian coordinates $i\in \lbrace x,y \rbrace$. 
We use an Einstein summation convention to sum over contracted indices i.e.
\begin{equation}
    a^i b_i = a^x b_x + a^y b_y.
\end{equation}
For Sec.~\ref{sec:qhregime}, we require an additional type of index for the flat internal space, which we denote $a, b, c, ... = 1,2$. 
In that section we also require a covariant notation for the external space-time manifold, and use Greek letters $\mu,\nu = t,x,y$ to denote external space-time indices. 

In all sections, external spatial indices are raised and lowered with the flat space metric $\delta^i_j$, since we are working with linearized (time-dependent) perturbations around flat space. 
Where possible, we retain the distinction between upper and lower indices in order to emphasize the different meaning of the indices in the stress tensor and viscosity tensor. 
As an example of our notation, we translate the Navier Stokes equation from Ref.~\onlinecite{landau1987fluid} into our notation, 
\begin{equation}
\bal
(\partial_t + \bf{v\cdot \boldsymbol{\nabla}} (\rho{\bf v}) + {\boldsymbol{\nabla} \cdot \boldsymbol{\tau}}) = 0\\
\Longrightarrow (\partial_t + v^i \partial_i) (\rho v^j) + \partial_i \tau^i_j = 0
\eal
\end{equation}
Using conservation of mass $\partial_t \rho + \partial_j (\rho v^i) = 0$ and taking the momentum density to be $g_j = \rho \,\delta_j^k v^k$, we recover Eq.~\eqref{eq:ns}:
\begin{equation}\label{eq:nssup}
    \partial_t g_j + \partial_i(\tau^i_{\hphantom{i}j}+g_jv^i ) =0,
\end{equation}
Since the stress tensor $\tau^i_j$ is not symmetric for anisotropic systems, it is important to distinguish between its two indices. 
The first upper index refers to the normal vector to a fluid
parcel, while the second lower index refers to the direction of the internal force. 
In
order to avoid confusion between these indices, we retain the notational distinction
between upper and lower indices in the stress and viscosity tensors, despite working in (and perturbing around) flat space for much of the work. Furthermore, we note that to make contact with the hydrodynamics literature, our choice $g_j = \rho \,\delta_j^k v^k$ implies that the stress tensor $\tau^i_{\hphantom{i}j}$ does not include corrections due to internal spin current (i.e., it is not the Belinfante stress of Refs.~\cite{bradlyn2014low,rao2019hall}). As we discuss in Sec.~\ref{sec:intro} and Appendix~\ref{app:ambiguities}, knowledge of the microscopic degrees of freedom is necessary to perform the Belinfante symmetrization.

\section{Review of Anisotropic Viscosity}\label{app:viscreview}
In this section we give a more general review of the anisotropic Hall viscosity, summarizing the setup of Ref.~\cite{rao2019hall}. 
Without any rotational symmetry and in the absence of time reversal symmetry, the Hall viscosity tensor is generically expressed in terms of six coefficients, \begin{align} \label{eq:hallviscgeneral}
    \visc{(\eta^\mathrm{H})}{i}{j}{k}{\ell}
    &\equiv\frac{1}{2}\left(\visc{\eta}{i}{j}{k}{\ell}-\visc{\eta}{k}{\ell}{i}{j}\right) \nonumber\\
    &=\eta^\mathrm{H}\visc{(\sigma^z\wedge\sigma^x)}{i}{j}{k}{\ell}+\gamma\visc{(\sigma^z\wedge\epsilon)}{i}{j}{k}{\ell}  \nonumber \\
    & +\Theta\visc{(\sigma^x\wedge\epsilon)}{i}{j}{k}{\ell}\nonumber +\bar{\eta}^\mathrm{H}\visc{(\delta\wedge\epsilon)}{i}{j}{k}{\ell}+\bar{\gamma}\visc{(\delta\wedge\sigma^x)}{i}{j}{k}{\ell}\nonumber  \\ &+\bar{\Theta}\visc{(\sigma^z \wedge \delta)}{i}{j}{k}{\ell},
\end{align}
Now when we look at the viscous forces produced in the bulk by this Hall viscosity tensor, we see that the barred and unbarred coefficients contribute to the same component of the bulk viscous force. 
In particular we have that the viscous force density is controlled by the rank two ``Hall tensor"
\begin{align}
    f^{\mathrm{H},\eta}_\mathrm{j}&=\sum_{\substack{ij'\ell'\\ \ell k}}\frac{1}{2}\left(\epsilon^{j'\ell'}\visc{(\eta^\mathrm{H})}{i}{j'}{k}{\ell'}\right)\partial_i\partial_k(\epsilon_{j\ell}v^\ell) \\
    &\equiv\sum_{ik\ell}\eta_\mathrm{H}^{ik}\partial_i\partial_k(\epsilon_{j\ell}v^\ell)\nonumber.\\
\end{align}
with
\begin{align}
    \label{eq:halltensor}
    \eta_\mathrm{H}^{ij}&=\frac{1}{4}\sum_{k\ell}\epsilon^{k\ell}\left(\visc{\eta}{i}{k}{j}{\ell}+\visc{\eta}{j}{k}{i}{\ell}\right)\\
    &= (\eta^\mathrm{H}+\bar{\eta}^\mathrm{H})\delta^{ij}+(\gamma+\bar{\gamma})\sigma_z^{ij}+(\Theta+\bar{\Theta})\sigma_x^{ij} \nonumber,
\end{align}
The coefficient $\eta^\mathrm{H}$ is the usual isotropic Hall viscosity \cite{avron1987adiabatic}, the coefficient $\bar{\eta}^\mathrm{H}$ breaks angular momentum conservation and can appear in active (or anisotropic) systems, and the rest of the coefficients are explicitly anisotropic and appear when a system has less than threefold rotation symmetry. 

\subsection{Non-dissipative contact terms}
As mentioned in Sec.~\ref{sec:intro}, the difference $\eta^\mathrm{H}_\mathrm{diff}\equiv \eta^\mathrm{H}-\bar{\eta}^\mathrm{H}$ between the isotropic Hall viscosities does not enter into the bulk force, it can be shifted by adding a divergenceless ``contact" \cite{rao2019hall} term 
\beq
\label{eq:contactterm}\delta \tau^i_{\hphantom{i}j} = C_0\partial^{*\; i} v_j
\eeq 
to the bulk stress tensor. 
From the lens of the viscosity tensor, the individual coefficients get shifted as
\beq
\bal
\eta^H \rightarrow \eta^H + C_0/2\\
\bar{\eta}^H \rightarrow \bar{\eta}^H - C_0/2, 
\eal
\eeq 
We note here that a more general expression of the contact term
\begin{equation}
    \delta\tau^i_{\hphantom{i}j}=\sum_{k\ell}\epsilon^{ik}C_{j\ell}\partial_k v^\ell,\label{eq:divfreecontactgeneral}
\end{equation}
with the more general form of the coefficient $C_{j\ell}$ now as a symmetric rank two tensor
\begin{equation}
\label{eq:contacttermtensor}
    C_{j\ell}=C_0\delta_{j\ell}+C_x \sigma^x_{j\ell}+C_z\sigma^z_{j\ell},
\end{equation}
In addition to the described effect of $C_0$, this provides the effect of shifting the difference between all barred and unbarred viscosities, and individually shifting the other viscosities as 
\begin{eqnarray}
\gamma&\rightarrow \gamma+C_z/2   \;\;\;\;\;\;\bar{\gamma}&\rightarrow \bar{\gamma}-C_z/2 \\
\Theta&\rightarrow \Theta+C_x/2  \;\;\;\;\;\;\bar{\Theta}&\rightarrow \bar{\Theta}-C_x/2.
\end{eqnarray}
We can continue viewing the contact terms as viscosities by looking at the boundary force provided by the contact term $C_0$, for example:
\beq
{\bf f}^{(C_0 ,\, \text{bdry})} = C_0 \left[  \left(\partial_{\bf s} v_{\bf n} + \frac{v_{\bf s}}{R} \right) \hat{\bf n} + \left(\partial_{\bf s} v_{\bf s} - \frac{v_{\bf n}}{R} \right) \hat{\bf s} \right].
\eeq
In the viewpoint that the contact term is a proxy for modified stress boundary conditions, with the above expression dictating the stress at the boundary.
\subsection{Dissipative viscosities \& contact term}\label{app:viscreview_diss}
With higher than twofold rotational symmetry\footnote{Technically for fourfold symmetric systems there are two independent shear viscosities--this detail will not be relevant for our analysis} the dissipative viscosity tensor for a fluid can be parametrized as
\begin{align} \label{eq:dissipativevisc}
    \visc{(\eta^\mathrm{D})}{i}{j}{k}{\ell}
    &\equiv\frac{1}{2}\left(\visc{\eta}{i}{j}{k}{\ell}+\visc{\eta}{k}{\ell}{i}{j}\right) \nonumber\\
    &=\eta^\mathrm{sh}\visc{(\sigma^x\odot\sigma^x+\sigma^z\odot\sigma^z)}{i}{j}{k}{\ell}+\eta^\mathrm{R}\visc{(\epsilon\odot\epsilon)}{i}{j}{k}{\ell}  \nonumber \\
    & +\eta^\mathrm{RC}\visc{(\delta\odot\epsilon)}{i}{j}{k}{\ell}+ \zeta\visc{(\delta\odot\delta)}{i}{j}{k}{\ell}\nonumber,
    \end{align}
The familiar bulk viscosity $\zeta$ and shear viscosity $\eta^\mathrm{sh}$ provide frictional forces in response to dynamic dilatations and volume-preserving shears, respectively. 
The rotational or vortex viscosity $\eta^\mathrm{R}$ breaks angular momentum conservation (analogous to $\bar{\eta}^\mathrm{H}$) and provides local resistive torques in response to vorticity. 
Lastly, $\eta^\mathrm{RC}$ is another dissipative viscosity that breaks angular momentum conservation For an incompressible fluid, $\eta^{\mathrm{RC}}$ and $\bar{\eta}^H$ provide the same stress both in the bulk and on the boundary, and so in our analysis we can set $\eta^\mathrm{RC}=0$ without loss of generality\footnote{We note there are other constraints on this coefficient $\eta^\mathrm{RC}$ considered in Ref.~\cite{monteiro2021hamiltonian}}.
In addition to the non-dissipative contact terms, there is another contact term that plays a similar role except for dissipative viscosities, and amounts to considering an antisymmetric piece of the tensor $C_{ij}$ in Eq.~\eqref{eq:contacttermtensor}. 
Explicitly this contact term is 
\begin{equation}
    \delta\tau^i_{\hphantom{i}j}= C_{\mathrm{dis}} 
    \sum_{k\ell}\epsilon^{ik}\epsilon_{j\ell}\partial_k v^\ell,\label{eq:divfreecontact}
\end{equation}
Similar to the non-dissipative case, the bulk dissipative forces only depend on the linear combination. 
This contact term shifts three viscosities when added in this case, 
\beq
\bal
\label{eq:Dshifts}
\eta^\mathrm{sh} &\rightarrow \eta^\mathrm{sh} - C_\mathrm{dis}/2\\
\eta^\mathrm{R} &\rightarrow \eta^\mathrm{R} + C_{\mathrm{dis}}/2\\
\zeta &\rightarrow \zeta + C_{\mathrm{dis}}/2
.\eal
\eeq
For the case of an incompressible fluid with $\zeta=0$, the contact term shifts the difference $\eta^\mathrm{dis}_\mathrm{diff}\equiv\eta^\mathrm{R} - \eta^\mathrm{sh}$, which is the case considered in the main text. 
We also note that it appears from the above that the contact term can generate a nonzero bulk viscosity for incompressible fluid with $\zeta=0$. 
In practice, however, this is unobservable as the dynamic constraint $\nabla \cdot {\bf v} = 0$ for an incompressible fluid prevents the bulk viscosity from contributing to the stress tensor.
For the threefold or higher rotationally symmetric case we consider in the main text,  the dissipative viscous force on the boundary is
\beq
\bal
\label{eq:dissbf}
{\bf f}^\mathrm{dis}&=  \left[\left(\eta^\mathrm{dis}_\mathrm{tot} + \eta^\mathrm{dis}_\mathrm{diff}\right) \partial_{\bf n} v_{\bf{n}} \right] {\bf \hat{n}} \\
&+ \left[\eta^\mathrm{dis}_\mathrm{tot} \omega + (\eta^\mathrm{dis}_\mathrm{tot}+\eta^\mathrm{dis}_\mathrm{diff}) \left(\partial_{\bf n} v_{\bf s} - \frac{v_{\bf s}}{R}\right) \right]{\bf \hat{s}}.
\eal
\eeq
Just as in the non-dissipative case, the boundary force depends not only on the bulk hydrodynamic observable $\eta^\mathrm{dis}_\mathrm{tot}$, but also on the difference $\eta^\mathrm{dis}_\mathrm{diff}$.

\subsection{Contact terms as ``magnetization stress"}

Let us consider the stress due to the combined dissipative and nondissipative contact terms,
\begin{align}
\delta \tau^i_{\hphantom{i}j} &=  \epsilon^{ik} (C_{jl}+C_\mathrm{dis}\epsilon_{jl})\partial_k v^l \\
&= -\epsilon^{ki}\partial_k\left[ (C_{jl}+C_\mathrm{dis}\epsilon_{jl}) v^l\right] \\
&= \epsilon^{kn}\partial_k m^i_{\hphantom{i}jn},
\end{align} 
where we have introduced the tensor
\begin{equation}
    m^i_{\hphantom{i}jn}=\delta^i_n\left[ (C_{jl}+C_\mathrm{dis}\epsilon_{jl}) v^l\right].
\end{equation}
We see then that the contact stress $\delta \tau^i_{\hphantom{i}j}$ is the curl of the vector of tensors $m^i_{\hphantom{i}jn}$. 
In analogy with electrodynamics, we can view $m^i_{\hphantom{i}jn}$ as a magnetization, and hence $\delta \tau^i_{\hphantom{i}j}$ is a magnetization stress (by analogy with magnetization currents).

Since $\delta \tau^i_{\hphantom{i}j}$ is divergenceless, in a uniform bulk it can only provide physical effects at system boundaries, consistent with the interpretation in terms of a magnetization. 
Conceptually the magnetization stresses are very similar to the energy magnetization currents in Refs.~\cite{halperin1993theory,bradlyn2014low}.

The density of power dissipated in an incompressible viscous fluid (with the same symmetry considerations as in the previous section) 
Microscopic considerations beyond hydrodynamics might modify this statement.

\subsection{Stress boundary conditions}
We detail the modified version of the no-stress boundary condition, relevant for the free surface fluid problem we consider later on, 
\beq
\hat{n}_i \tau^{i}_{\hphantom{i}j} = 0.
\eeq
For a fluid with pressure $p$, we have the following conditions for the normal and tangential forces on the boundary:
\beq
\bal
\label{eq:nostressbcs_sm}
\hat{n}_i\hat{n}^j \tau^{i}_{\hphantom{i}j} =-p + \left(\eta^\mathrm{H}_\mathrm{tot} + \eta^\mathrm{H}_\mathrm{diff}\right) \left( \partial_{\bf s} v_{\bf n} + \frac{v_{\bf s}}{R} \right) + \eta^\mathrm{H}_\mathrm{tot} \omega + \left(\eta^\mathrm{dis}_\mathrm{tot} + \eta^\mathrm{dis}_\mathrm{diff}\right) \partial_{\bf n} v_{\bf{n}} &=0,\\
\hat{n}_i\hat{s}^j \tau^{i}_{\hphantom{i}j} =\left(\eta^\mathrm{H}_\mathrm{tot} + \eta^\mathrm{H}_\mathrm{diff}\right) \left( \partial_{\bf s} v_{\bf s} - \frac{v_{\bf n}}{R} \right) + \eta^\mathrm{dis}_\mathrm{tot} \omega + (\eta^\mathrm{dis}_\mathrm{tot}+\eta^\mathrm{dis}_\mathrm{diff}) \left(\partial_{\bf n} v_{\bf s} - \frac{v_{\bf s}}{R}\right) &=0.
\eal
\eeq

\section{Power dissipation \& the total dissipative viscosity}\label{app:power}

In this section, we will explore the implications of the viscous ambiguity on expressions for power dissipation and entropy production in a fluid. 
We will follow the logic of Ref.~\cite{landau1987fluid}, generalizing where necessary to allow for anisotropic and nondissipative viscosity. 
We will begin by considering energy dissipation in an incompressible, anisotropic fluid, and then generalize to incompressible fluids as well. 
For illustrative purposes, we will consider only fluids with particle rotation rate $\Omega=0$; the case of $\Omega\neq0$ can be treated by similar methods, replacing $\nabla\times\mathbf{v}\rightarrow \nabla\times\mathbf{v}-2\Omega$.

\subsection{Incompressible fluids}
 As a starting point, let us consider an incompressible fluid with nonzero $\eta^\mathrm{sh},\eta^\mathrm{R},\eta^\mathrm{H},$ and $\bar{\eta}^\mathrm{H}$. 
 The equations of motion for this fluid are

\begin{align}
\frac{\partial\rho}{\partial t} + \mathbf{v}\cdot\nabla\rho &= 0 \\
\rho\frac{\partial v^j}{\partial t} + \rho v^i\partial_i v^j &= -\partial_j p +\partial_i\left(\visc{\eta}{i}{j}{k}{\ell}\partial_k v^\ell\right)\label{eq:nssupp}.
\end{align} 
The first equation expresses conservation of mass, while the second equation is the Navier-Stokes equation with general viscosity tensor. 
We have left the viscous force written as $\partial_i\left(\visc{\eta}{i}{j}{k}{\ell} \partial_k v^\ell\right)$ for bookkeeping purposes - as discussed in Sec.~\ref{sec:intro}
\begin{equation}
\partial_i\left(\visc{\eta}{i}{j}{k}{\ell} \partial_k v^\ell\right) = (\eta^\mathrm{sh} + \eta^\mathrm{R})\nabla^2 v_j + (\eta^\mathrm{H}+\bar{\eta^\mathrm{H}})\nabla^2v^{*}_{j}.
\end{equation}
We can use the equations of motion to derive an expression for the rate of change of kinetic energy $E_\mathrm{k}$ of the fluid in a volume $V$. 
First, note that
\begin{equation}
E_\mathrm{k} = \int dV \frac{1}{2}\rho v^2.
\end{equation}
Taking the time derivative, we have
\begin{align}
\dot{E}_\mathrm{k}&= \int dV \left[\frac{1}{2}v^2\frac{\partial \rho}{\partial t} + 
\delta_{ik} v^i\rho\frac{\partial v^k}{\partial t}\right] \\
&=\int dV \left[ -\frac{1}{2}v^2v^i\partial_i \rho -\rho v^iv^j\partial_jv_i - v^i\partial_i p +v^j\partial_i(\visc{\eta}{i}{j}{k}{\ell}\partial_k v^\ell) \right].
\end{align}
Using the incompressibility of the fluid, we can rewrite the first three terms as a total divergence to obtain
\begin{equation}
\dot{E}_\mathrm{k}= \int dV \left[-\nabla\cdot\left(\frac{1}{2}\rho v^2\mathbf{v} + \mathbf{v}p\right) +v^j\partial_i(\visc{\eta}{i}{j}{k}{\ell}\partial_k v^\ell)\right]\label{eq:edotintermediate}.
\end{equation}
We would now like to perform a partial integration on the second term. 
There is an ambiguity in how we do this, however, due to the viscous redundancy. 
In particular, we can write
\begin{equation}
\partial_i(\visc{\eta}{i}{j}{k}{\ell}\partial_kv^\ell) = \partial_i(\visc{\eta}{i}{j}{k}{\ell}\partial_kv^\ell + \epsilon^{ik}\tilde{C}_{j\ell}\partial_kv^\ell),\label{eq:viscantideriv}
\end{equation}
where $\tilde{C}_{j\ell}=C_{j\ell} + C_\mathrm{dis}\epsilon_{j\ell}$, and $C_{j\ell}$ was defined in Eq.~\eqref{eq:contacttermtensor}. 
We see that $\tilde{C}_{j\ell}$ parametrizes the fact that the antiderivative of $\partial_i(\visc{\eta}{i}{j}{k}{\ell}\partial_kv^\ell)$ is determined only up to a divergenceless term, and is equivalent to our original viscous ambiguity. 
Inserting Eq.~\eqref{eq:viscantideriv} into Eq.~\eqref{eq:edotintermediate} and integrating by parts yields
\begin{align}
\dot{E}_\mathrm{k}&= \int dV \left[-\nabla\cdot\left(\frac{1}{2}\rho v^2\mathbf{v} + \mathbf{v}p - v^j(\visc{\eta}{i}{j}{k}{\ell}+\epsilon^{ik}\tilde{C}_{j\ell})\partial_kv^\ell \right) -\partial_i v^j(\visc{\eta}{i}{j}{k}{\ell}+\epsilon^{ik}\tilde{C}_{j\ell})\partial_kv^\ell\right] \\
&=-\oint da_i\left[v_i\left(\frac{1}{2}\rho v^2 +p\right) - v^j(\visc{\eta}{i}{j}{k}{\ell} + \epsilon^{ik}\tilde{C}_{j\ell})\partial_kv^\ell\right] - \int dV\left[\partial_i v^j(\visc{\eta}{i}{j}{k}{\ell}+\epsilon^{ik}\tilde{C}_{j\ell})v^j\partial_kv^\ell\right] \\
&\equiv -\oint d\mathbf{a}\cdot \mathbf{j}^\mathrm{KE} - \int{dV} W,
\end{align}
where we have defined the kinetic energy flux density
\begin{equation}
j^{\mathrm{KE},i} = v^i\left(\frac{1}{2}\rho v^2 + p\right) - v^j(\visc{\eta}{i}{j}{k}{\ell}+\epsilon^{ik}\tilde{C}_{j\ell})\partial_kv^\ell,\label{eq:keflux}
\end{equation}
and the local heating rate
\begin{equation}
W = (\visc{\eta}{i}{j}{k}{\ell}+\epsilon^{ik}\tilde{C}_{j\ell})\partial_iv^j\partial_kv^\ell\label{eq:heating}.
\end{equation}
We see that the ambiguity $\tilde{C}_{j\ell}$ in defining the divergence-free part of the stress tensor results in an ambiguity in the way we separate kinetic energy loss into flux out of a volume (captured by $\mathbf{j}^\mathrm{KE}$) and local dissipation (captured by $W$). 
Writing out $W$ for the case at hand with $\tilde{C}_{j\ell}=C_0\delta_{j\ell}+C_\mathrm{dis}\epsilon_{j\ell}$, we find that
\begin{align}
W &= \eta^\mathrm{sh}\left(\partial_iv^j+\partial_jv^i\right)\partial_iv^j + \eta^\mathrm{R}\left(\partial_iv^j-\partial_jv^i\right)\partial_iv^j - C_\mathrm{dis}\partial_iv^j\partial_jv^i \\
&=\eta^\mathrm{sh}\left(\partial_iv^j+\partial_jv^i\right)\partial_iv^j + \eta^\mathrm{R}\left(\partial_iv^j-\partial_jv^i\right)\partial_iv^j - \frac{C_\mathrm{dis}}{2}\left(\partial_jv^i+\partial_iv^j\right)\partial_iv^j -\frac{C_\mathrm{dis}}{2}\left(\partial_jv^i-\partial_iv^j\right)\partial_iv^j \\
&=\left(\eta^\mathrm{sh}-\frac{C_\mathrm{dis}}{2}\right)\left(\partial_iv^j+\partial_jv^i\right)\partial_iv^j + \left(\eta^\mathrm{R}+\frac{C_\mathrm{dis}}{2}\right)\left(\partial_iv^j-\partial_jv^i\right)\partial_iv^j \\
&=\frac{1}{2}\left(\eta^\mathrm{sh}-\frac{C_\mathrm{dis}}{2}\right)\left(\partial_iv^j+\partial_jv^i\right)^2 + \left(\eta^\mathrm{R}+\frac{C_\mathrm{dis}}{2}\right)(\nabla\times\mathbf{v})^2.
\end{align}
This is precisely the local dissipation rate predicted from classical fluid dynamics with an effective shear viscosity $\left(\eta^\mathrm{sh}-\frac{C_\mathrm{dis}}{2}\right)$, and an effective rotational viscosity $\left(\eta^\mathrm{R}+\frac{C_\mathrm{dis}}{2}\right)$. 

In order to unambiguously relate the local heating rate to the viscosity coefficients $\eta^\mathrm{sh}$ and $\eta^\mathrm{R}$, the energy flux $\mathbf{j}^\mathrm{KE}$ needs to be determined. 
Given a microscopic model for a fluid, such as a continuum field theory, the energy flux may be uniquely specifiable via a (minimal) coupling of the fluid to background fields\cite{bradlyn2014low,cook2019electron,huang2022hydrodynamic}. 
Such a procedure would fix a value for $\tilde{C}_{j\ell}$, and hence fix a relationship between $W$ and the viscosity coefficients\cite{cook2021viscometry}. 
However, absent such a microscopic model, an experimental method for determining $\mathbf{j}^\mathrm{KE}$ independently of $W$ is needed to extract the viscosity coefficients from local heating.

One might worry that changing the definition of $\mathbf{j}^\mathrm{KE}$ may have implications for the stability of the fluid. 
In the standard approach, bulk stability of a fluid necessitates that the $\dot{E}_\mathrm{k} \le 0$ when integrated over the entire fluid, assuming that the velocity of the fluid goes to zero at infinity (or at the boundary). 
In this case, we have that the boundary integral of $\mathbf{j}^\mathrm{KE}$ goes to zero, and
\begin{equation}
\dot{E}_\mathrm{kin}^\mathrm{tot} = -\int dV W = \int dV (\eta^\mathrm{sh}+\eta^\mathrm{R})\mathbf{v}\cdot\nabla^2\mathbf{v}.
\end{equation}
where we used the vanishing of the velocity at the boundary of the volume to integrate by parts. 
Since $\mathbf{v}\cdot\nabla^2\mathbf{v}$ is negative definite, we see that the fluid is stable provided $(\eta^\mathrm{sh}+\eta^\mathrm{R}) > 0$, as stated below Eq.~\eqref{eq:distotdef}. 
Crucially, we notice that our choice of $\tilde{C}_{j\ell}$ in defining the energy flux does not play a role.

Note also that although the Hall viscosities and $C_0$ do not contribute to local heating, they \emph{do} contribute to the kinetic energy flux $\mathbf{j}^\mathrm{KE}$. 
In fact, by direct substitution we have
\begin{align}
v^j\left(\visc{(\eta^\mathrm{H})}{i}{j}{k}{\ell} + C_0\epsilon^{ik}\delta_{j\ell}\right)\partial_kv^\ell = \eta^\mathrm{H}\left[(\mathbf{v}\cdot\nabla) v^{*,i}+(\mathbf{v}\times\nabla)v^i\right] + \bar{\eta}^\mathrm{H}v^i\nabla\times\mathbf{v} + C_0v_\ell\partial^{*,i}v^\ell.
\end{align}
This does not vanish. 
This means that the non-dissipative viscosity contributes to kinetic energy flux, even though it does not contribute to local heating. 
Focusing on the contribution from the Hall viscous ambiguity, however, we have that
\begin{equation}
C_0v_\ell\partial^{*,i}v^\ell = \frac{C_0}{2}\partial^{*,i} v^2
\end{equation}
is a total (exterior) derivative; for any nonsingular velocity field, the integral of this term around any closed curve is zero. 
Thus, the Hall viscous ambiguity has no impact on the integrated kinetic energy flux. 
In this sense, $C_0$ determines the definition of the energy magnetization current familiar in quantum Hall systems\cite{cooper1997thermoelectric}.
\subsection{Compressible fluids}

It is natural to ask what happens to these considerations when we look at compressible fluids. 
To do so, we must allow $\nabla\cdot\mathbf{v}\neq 0$, and so generically the bulk viscosity $\zeta\neq 0$ and the additional torque $\eta^\mathrm{RC}\neq 0$. 
For a compressible fluid, the conservation of mass takes the more general form
\begin{align}
\frac{\partial\rho}{\partial t} + \nabla\cdot(\rho\mathbf{v}) &= 0,
\end{align}
which ensures that the Navier-Stokes equation Eq.~\eqref{eq:nssupp} remains unchanged (aside from the inclusion of the more general viscosity tensor). 
We can now look at the total energy density for our compressible fluid,
\begin{equation}
\epsilon = \frac{1}{2}\rho v^2 + \rho u,
\end{equation}
where $u$ is the thermodynamic internal energy per unit mass of the fluid. 
We would like to examine the rate of change $\partial\epsilon/\partial t$. 
For concreteness, we will consider a fluid where the dynamics conserve energy locally, such that we expect 
\begin{equation}
\frac{\partial \epsilon}{\partial t} + \nabla\cdot \mathbf{j}^\mathrm{E} = 0\label{eq:energycons},
\end{equation}
for some definition of the total energy flux $\mathbf{j}^\mathrm{E}$ to be determined.

To proceed, it is useful to recall a few thermodynamic identities. 
From the first law of thermodynamics, we have that when the number of fluid particles is conserved
\begin{equation}
du = Tds -pd\frac{1}{\rho} = Tds + \frac{p}{\rho^2}d\rho\label{eq:firstlaw}.
\end{equation}
To separate out heating from work done by the fluid, it will also be useful to introduce the enthalpy per unit mass
\begin{equation}
w=u+\frac{p}{\rho}\label{eq:enthalpy},
\end{equation}
whose differential change is given by
\begin{equation}
dw=Tds+\frac{1}{\rho}dp\label{eq:enthalpydiff}.
\end{equation}
Using these relations, We can compute
\begin{equation}
\frac{\partial\epsilon}{\partial t} = (\frac{1}{2}v^2+u)\frac{\partial\rho}{\partial t} + \frac{1}{2}\rho\frac{\partial{v^2}}{\partial t} + \rho\frac{\partial u}{\partial t}\label{eq:etotderiv}.
\end{equation}
From the first law Eq.~\eqref{eq:firstlaw} we have
\begin{equation}
\rho\frac{\partial u}{\partial t}  = \rho T \frac{\partial s}{\partial t} +\frac{p}{\rho}\frac{\partial \rho}{\partial t}.
\end{equation}
We can insert this into Eq.~\eqref{eq:etotderiv} and use our definition of $w$ to find
\begin{equation}
\frac{\partial\epsilon}{\partial t} = \left(\frac{1}{2}v^2+w\right)\frac{\partial\rho}{\partial t}+\frac{1}{2}\rho\frac{\partial v^2}{\partial t} + \rho T \frac{\partial s}{\partial t}.
\end{equation}
Using next conservation of mass and the Navier-Stokes equation yields
\begin{equation}
\frac{\partial\epsilon}{\partial t} = -\left(\frac{1}{2}v^2+w\right)\partial_i(\rho v^i)+\rho T \frac{\partial s}{\partial t}-\rho v_jv^i\partial_i v^j -v^j\partial_j p +v^j\partial_i(\visc{\eta}{i}{j}{k}{\ell}\partial_kv^\ell).\label{eq:etotintermediate}
\end{equation}
We can now eliminate the gradient of the pressure using Eq.~\eqref{eq:enthalpydiff} to write
\begin{equation}
\nabla p = \rho \nabla w -\rho T \nabla s.\label{eq:pgrad}
\end{equation}
Additionally, note that
\begin{equation}
\frac{1}{2}v^2\partial_i(\rho v^i) + \rho v_jv^i\partial_i v^j =\partial_i (\frac{1}{2}\rho v^2 v^i).\label{eq:kgrad}
\end{equation}
Inserting Eqs.~\eqref{eq:pgrad} and \eqref{eq:kgrad} into Eq.~\eqref{eq:etotintermediate} yields
\begin{equation}
\frac{\partial\epsilon}{\partial t} = -\partial_i\left(\frac{1}{2}\rho v^2v^i + \rho wv^i\right)+\rho T\left(\frac{\partial s}{\partial t} + v^i\partial_i s\right) +v^j\partial_i(\visc{\eta}{i}{j}{k}{\ell}\partial_kv^\ell).
\end{equation}
Up to now, we have followed the usual derivation (of, e.~g., Ref.~\cite{landau1987fluid}) quite closely. 
To make further progress, we will introduce the general antiderivative Eq.~\eqref{eq:viscantideriv} just as before, and integrate by parts. 
We find then that
\begin{align}
\frac{\partial\epsilon}{\partial t} &= -\partial_i\left(\frac{1}{2}\rho v^2v^i + \rho wv^i -v^j(\visc{\eta}{i}{j}{k}{\ell}+\epsilon^{ik}\tilde{C}_{j\ell})\partial_kv^\ell\right)+\rho T\left(\frac{\partial s}{\partial t} + v^i\partial_i s\right) +\partial_iv^j(\visc{\eta}{i}{j}{k}{\ell}+\epsilon^{ik}\tilde{C}_{j\ell})\partial_kv^\ell \\
&=-\partial_i\left(j^{\mathrm{KE},i} +\rho u v^i\right) + \rho T\left(\frac{\partial s}{\partial t} + v^i\partial_is - \frac{W}{\rho T}\right)\label{eq:dedtalmostfinal},
\end{align}
where we have reintroduced the kinetic energy flux $\mathbf{j}^\mathrm{KE}$ and the local heating $W$ from Eqs.~\eqref{eq:keflux} and \eqref{eq:heating} respectively, allowing here for a more general $\visc{\eta}{i}{j}{k}{\ell}$. 
To allow for thermal conductivity in the fluid, we can also add and subtract to Eq.~\eqref{eq:dedtalmostfinal} the energy flux from thermal conductivity
\begin{equation}
q = \partial_i\lp\kappa^{ij}\partial_j T\rp.
\end{equation}
We can then write
\begin{equation}
\frac{\partial \epsilon}{\partial t} = -\partial_i({j}^{\mathrm{E},i} ) +\rho T\left(\frac{\partial s}{\partial t} + v^i\partial_is - \frac{W}{\rho T} -\frac{1}{\rho T}\partial_i(\kappa^{ij}\partial_j T)\right).
\end{equation}
where we have introduced the total energy flux
\begin{equation}
{j}^{\mathrm{E},i}=j^{\mathrm{KE},i} + \rho u v^i -\kappa^{ij}\partial_j T\label{eq:efluxfinal},
\end{equation}
accounting for the flow of kinetic energy in the first term, internal energy in the second term, and heat in the third term. 
Using the conservation equation Eq.~\eqref{eq:energycons}, we have that
\begin{align}
\frac{\partial\epsilon}{\partial t} + \nabla\cdot\mathbf{j}^\mathrm{E}&=0 \\
\frac{\partial s}{\partial t} +\mathbf{v}\cdot\nabla s  &=\frac{1}{\rho T}(W+q).
\end{align}
The first term is a restatement of energy conservation, although now the energy flux is defined by Eq.~\eqref{eq:efluxfinal}; it depends on the bulk viscous ambiguity $\tilde{C}_{j\ell}$ via Eq.~\eqref{eq:keflux}. 
As before, this means we can only uniquely express the heating $W$ in terms of the bulk viscosity coefficients once we fix a definition for the energy flux, either via an experiment, or via microscopic considerations such as coupling to background geometric fields. 

We see also that the local heating $W$ indeed controls the rate of change of entropy due to fluid friction. 
Inserting our explicit form of the viscosity tensor into Eq.~\eqref{eq:heating}, we find that
\begin{equation}
W=\frac{1}{2}\left(\eta^\mathrm{sh}-\frac{C_\mathrm{dis}}{2}\right)\left(\partial_iv^j+\partial_jv^i-\delta_i^j\nabla\cdot\mathbf{v}\right)^2 + \left(\eta^\mathrm{R}+\frac{C_\mathrm{dis}}{2}\right)(\nabla\times\mathbf{v})^2 + \left(\zeta +\frac{C_\mathrm{dis}}{2}\right)(\nabla\cdot\mathbf{v})^2 +2\eta^\mathrm{RC}(\nabla\cdot \mathbf{v})(\nabla\times\mathbf{v}).
\end{equation}
Note that the presence of nonzero $\eta^\mathrm{RC}$ entangles the $\eta^\mathrm{R}$ and $\zeta$ contributions to the local heating. in fact, we can write
\begin{equation}
\frac{1}{2}\left(\eta^\mathrm{sh}-\frac{C_\mathrm{dis}}{2}\right)\left(\partial_iv^j+\partial^jv_i-\delta^i_j\nabla\cdot\mathbf{v}\right)^2 + \begin{pmatrix}
\nabla\times\mathbf{v} & \nabla\cdot\mathbf{v}\end{pmatrix}\begin{pmatrix}
\eta^\mathrm{R} + \frac{C_\mathrm{dis}}{2} & \eta^\mathrm{RC} \\
\eta^\mathrm{RC} & \zeta + \frac{C_\mathrm{dis}}{2}
\end{pmatrix}\begin{pmatrix}
\nabla\times\mathbf{v} \\
\nabla\cdot\mathbf{v}\end{pmatrix}.
\end{equation}

To verify that the second law of thermodynamics is satisfied, we can examine the rate of change of the total entropy of the fluid,
\begin{align}
\dot{S} &= \int dV \frac{\partial \rho s}{\partial t} \\
& = \int dV \rho\frac{\partial s}{\partial t} - s\nabla\cdot\mathbf{\rho v} \\
&=-\int dV \nabla\cdot{\rho v s} +\frac{1}{T}(W+q).
\end{align}
The first term is the total entropy flux out to infinity, while the second term describes local entropy generation. 
Since we are integrating over the whole fluid, and if we assume that gradients of the fluid velocity vanish at infinity, the first term integrates to zero. 
The second law of thermodynamics then demands that
\begin{equation}
\int dV \frac{1}{T}(W+q) \geq 0.
\end{equation}
For this to be true for any flow, it must be the case that each quadratic form appearing in the integrand must be non-negative. 
This means, in particular, that
\begin{align}
\eta^\mathrm{sh} &\geq \frac{C_\mathrm{dis}}{2}\label{eq:shearconst} \\
\zeta + \eta^\mathrm{R} &\geq -C_\mathrm{dis} \\
\left(\eta^\mathrm{R}+\frac{C_\mathrm{dis}}{2}\right)\left(\zeta+\frac{C_\mathrm{dis}}{2}\right) &\geq \left(\eta^\mathrm{RC}\right)^2\label{eq:rcconst} \\
\mathrm{tr}(\kappa) &\geq 0 \\
\mathrm{det}(\kappa^s) &\geq0,
\end{align}
where $\kappa^s_{ij}=1/2(\kappa_{ij}+\kappa_{ji})$ is the symmetric part of the thermal conductivity. 
Note that Eqs.~\eqref{eq:shearconst}--\eqref{eq:rcconst} imply that
\begin{align}
\eta^\mathrm{R} + \eta^\mathrm{sh} &\geq 0 \\
\zeta+\eta^\mathrm{sh} &\geq 0,
\end{align}
independent of the choice of energy flux (i.~e. independent of $\tilde{C}_{ij}$). 
These same constraints were also derived in Ref.~\cite{cook2019electron} by considering damping of sound waves in the bulk. 
We see, additionally that a nonzero $\eta^{\mathrm{RC}}$ places an additional nontrivial constraint on the energy flux. 
Note also that when $\zeta=0$, these coincide with the stability conditions derived in Sec.~\ref{sec:waves} from surface wave stability.

To summarize, we see that for a general fluid, the bulk equations of motion do not uniquely specify the stress tensor, and hence do not uniquely specify the viscosity tensor. 
The ambiguity is due to divergenceless contributions to the stress tensor that do not enter into the bulk equations of motion. 
From the point of view of energy flow, this results in an ambiguity in defining the energy flux in the system. 
We have shown that the local heating rate depends not just on the viscosity coefficients in the bulk, but also on the choice of energy flux.
We emphasize that, given a microscopic description of the fluid dynamics, or alternatively a field theory description of the fluid coupled to background geometric fields, it is possible to define the energy flux and hence compute $\tilde{C}_{ij}$ from first principles--in most textbook descriptions, $\tilde{C}_{ij}=0$.

\section{Stress ambiguities, contact terms \& internal angular momentum}\label{app:ambiguities}

In this section we aim to emphasize that unless there is a procedure to determine the antisymmetric stress, or in particular measure the internal angular momentum $L_\mathrm{int}$ in the bulk of a system, the contact terms and viscous ambiguity must be present. 
Ambiguities in the stress tensor appear in other contexts, most notably through the Irving-Kirkwood formalism for treating the stress tensor for interacting systems \cite{irving1950statistical,fruchart2022odd,bradlyn2012kubo}. 
In that picture, for a two-body interaction, one must define a path length between interacting constituents, and there is an inherent ambiguity in choosing this path. 
In Ref.~\cite{bradlyn2012kubo}, for example, the authors choose the geodesic distance between particles, which is the standard approach and agrees with the stress tensor in theories of gravity. 

We can shift between various choices of this path by adding divergenceless pieces to the stress tensor, which are very similar to the contact terms mentioned in the present work and in Ref.~\cite{rao2019hall}. 
The main difference is that the divergenceless terms arising through the Irving-Kirkwood formalism are proportional to gradients in the fluid density and therefore are not viscous, whereas the contact terms of this work are manifestly viscous contributions to the stress. 
In the case where interactions between particles are spin-orbit coupled, such as in the active fluids we consider in the present work, the pertinent ambiguity relates to the antisymmetric part of the stress tensor and specifically the internal angular momentum generator $L_\mathrm{int}$.

In this work we are able to tie the contact terms to physical observables and resolve the viscous ambiguity, showing for example that different choices of the stress tensor yield different dispersions for surface waves. 
We can then revisit situations where the internal angular momentum $L_\mathrm{int}$ is unclear (such as in Ref.~\cite{soni2019odd} where it is assumed to be constant), and from purely boundary information, learn more about the true form of the bulk internal angular momentum. 

\subsection{Microscopic example of viscous ambiguity}\label{eq:app_massaniso}
As a further example of a system where the viscous ambiguity is present, consider a quantum Hall system with band mass anisotropy\cite{offertaler2019viscoelastic},
\begin{equation}
    \label{eq:bandmass}
    H_{AM} = \frac{1}{2} \tilde{m}_{ab} \pi^a \pi^b,  \;\; \curl{\vec{A}} = B\hat{z}.
\end{equation}
The inverse mass tensor is a symmetric and diagonalizable matrix so we choose a basis where
\begin{equation}\tilde{m}_{ab} = m^\delta \delta_{ab} + m^{\sigma^z} \sigma^z_{ab}.
\end{equation} 
The Hall viscosity can be calculated\cite{offertaler2019viscoelastic} to find
\begin{equation}
    \visc{(\eta^\mathrm{H})}{a}{b}{c}{d}= \frac{\hbar \rho}{4} \left[\delta^{a}_{d}\epsilon_b^e (m\tilde{m}_{ce}) + \delta_b^c \epsilon^e_d (m\tilde{m}_{ae}) \right].
\end{equation}
In Appendix~\ref{app:viscreview}, we introduced the Hall tensor Eq.~\eqref{eq:halltensor} as a useful device to parameterize the total viscosities that contribute to bulk forces; for this model, it is given by
\begin{equation}
\label{eq:visctensor}
    \eta^H_{ab} = \frac{\hbar\rho}{4}\left(\frac{1}{m} \tilde{m}_{ab}\right) = \frac{\hbar \rho}{4m} \left(m^\delta \delta_{ab} + m^{\sigma^z} \sigma^z_{ab}\right).
\end{equation}
Comparing with Eq.~\eqref{eq:halltensor}, we can identify the coefficients 
\begin{equation}
\begin{aligned}
    \eta^\mathrm{H}_\mathrm{tot} &=  \frac{\hbar\rho}{4m} m^\delta\\
    \gamma_\mathrm{tot} &= \frac{\hbar\rho}{4m}m^{\sigma^z}.
    \end{aligned}
\end{equation}
We can also read off the viscosity components from the full tensor Eq.~\eqref{eq:visctensor} and find which specific components the total viscosities are comprised of:
\begin{equation}
\begin{aligned}
    \eta^\mathrm{H}_\mathrm{tot}= \eta^\mathrm{H}\\
    \gamma_\mathrm{tot} = \gamma.
\end{aligned}
\end{equation}
The point we make is that this \textit{Hall tensor} matches that of another system, a 3D system with tilted field anisotropy projected to two dimensions, with Hamiltonian\cite{offertaler2019viscoelastic}
\begin{equation}
    H_{TF} = \frac{1}{2m} \pi^\mu \pi_\mu + \frac{1}{2} m\omega_0^2 z^2 \;\; \text{with}\;\; \curl{\vec{A}} = B_x \hat{x} + B_z \hat{z}.
\end{equation}
This model has a symmetric stress tensor as opposed to the band mass anisotropic system, and so while the Hall tensor for both systems are equivalent, the individual viscosity components differ. 
When the tilted-field system is projected to two dimensions, if we scale the density and out-of-plane magnetic field as $\rho^* = \rho(1-k^2l^2), B_z^* = B_z(1-k^2l_z^2)$, the system can be viewed as a two dimensional system with effective band mass anisotropy
\begin{equation}
    m^\mathrm{eff}_{ab} = m\begin{pmatrix}1-\frac{1}{2}k^2l^2 &0\\0& 1+\frac{1}{2}k^2l^2 
    \end{pmatrix}.
\end{equation}
The Hall tensor for the tilted field system then takes the form\cite{offertaler2019viscoelastic}
\begin{equation}
\eta^H_{ab} = \frac{\hbar\rho^*}{4}\left(\frac{1}{m} m^\mathrm{eff}_{ab}\right).
\end{equation}
The stress tensor is symmetric for the model Eq.~\eqref{eq:bandmass}, so this tells us that the total viscosities are entirely due to the following individual components $\eta^\mathrm{H}_\mathrm{tot} = \eta^\mathrm{H}$ and $\gamma_\mathrm{tot} = \bar{\gamma}$. 
Thus although the tilted field and anisotropic mass systems have identical $\eta^H_\mathrm{tot}$ and $\gamma_\mathrm{tot}$, they have opposite difference viscosities $\gamma-\bar{\gamma}$. 
Therefore the stress tensors for the two systems differ by a divergenceless contact term. 
This serves as a quantum example of the viscous ambiguity. 

\subsection{Resolution of ambiguity with internal angular momentum}
If the internal angular momentum $L_\mathrm{int}$ of a system is known, then the contact terms are fixed and the viscosities are unambiguous. 
For a quantum fluid, the procedure to correct the antisymmetric part of the stress with $L_\mathrm{int}$ is elaborated in detail in Ref.~\cite{rao2019hall}. 
Here, we summarize this construction for a classical fluid. 
We assume translational invariance so that we have conserved momentum density
\begin{equation}
    \partial_t{g_j({\bf r})} = -\partial_i \tau^i_{\hphantom{i}j}.
\end{equation}
We also consider total angular momentum conservation, where the total angular momentum is $L_\mathrm{tot} = L_\mathrm{orb} + L_\mathrm{int}$ and the orbital angular momentum is expressed from the kinetic momentum density as $L_\mathrm{orb} = \epsilon^i_j (x_i g^j)$. 
Angular momentum conservation is then stated as 
\begin{equation}
    \partial_t L_\mathrm{tot} = \partial_k M^{\mathrm{tot},k}.
\end{equation}
The tensor $M^{\mathrm{tot},k}$ parametrizes the flux of angular momentum, and is determined from microscopic considerations. 
This can also be decomposed into orbital and internal parts
\begin{equation}
M^{\mathrm{tot},k} = \epsilon^{i}_{\hphantom{i}j} (x_i \tau^{jk}) + M^{\mathrm{int},k}.
\end{equation}
Combined with the continuity equation, this leads to a constraint on the antisymmetric stress:
\begin{equation}
\label{eq:angmomconst}
    \partial_t L_\mathrm{int}({\bf r},t) = \epsilon^j_i\tau^i_{\hphantom{i}j} + \partial_k M^{\mathrm{int},k}.
\end{equation}
The equation above tells us that angular momentum conservation and knowledge of $L_\mathrm{int}$ and $M^{\mathrm{int},k}$ place a constraint on the antisymmetric stress of the system and fixes the contact terms. 
Furthermore, if were we to shift the angular momentum flux tensor, we could shift the antisymmetric stress. 
For an incompressible fluid, this shifts the difference between $\eta^\mathrm{sh}$ and $\eta^\mathrm{R}$, which is also the effect of the dissipative contact term. 
As a result, if we can measure not only the fluid velocity, but also the fluid angular momentum, then it is possible to uniquely determine the viscosity coefficients with bulk measurements. 
Otherwise, boundary measurements are necessary to experimentally fix the viscosity coefficients.

There are experimental settings where this type of resolution of the viscous ambiguity is feasible, namely nematic liquid crystals\cite{parodi} and hydrodynamic electron fluids with spin-orbit coupling\cite{denisov2022spin}. 
These settings provide natural bulk constitutive relations for internal angular momentum (the director for liquid crystal and spin for the electron fluid). 
We leave a more detailed experimental proposal for the measurement of individual viscosity coefficients and resolution of the ambiguity in these settings for future work. 

\subsection{Torsional Hall viscosity and viscous ambiguity}

In a system where torsion is treated as independent and can be non-minimally coupled to, a torsional Hall viscosity $\xi^\mathrm{H}$ \cite{hughesleighfradkin} can arise, which is due to the stress response of the following effective action,
\begin{equation}
    S_\mathrm{eff} = \xi^\mathrm{H} \int e^a \wedge T^b \eta_{ab}.
\end{equation}
where $e^a=e^a_\mu dx^\mu$ are the vielbeins introduced in Sec.~\ref{sec:qhregime}, and the torsion 2-form $T^a$ is given by $T^a = d{\bf e}^a + \omega^a_b \wedge {\bf e}^b$ with spin connection $\omega^a_b$. 
If the torsion is fixed, the spin connection is determined by the vielbeins. 
If we consider linear perturbations of the vielbein $\partial_t e^a_\mu = \partial_\mu v^a$, the (spatial) stress response from this action is given by
\begin{equation}
    T_{ij} = -\xi^\mathrm{H} \epsilon^{ik}  \partial_k v^j = -\xi^\mathrm{H} \partial^*_i v^j.
\end{equation}
We recognize the form of this (viscous) stress in Eq.~\eqref{eq:contactterm}; it matches the stress due to a non-dissipative contact term $C_0= -\xi^\mathrm{H}$. 
In short, the torsional Hall viscosity is a contact term, and produces a magnetization stress. 
It is therefore no surprise that upon symmetrization of the stress tensor (disallowing non-minimal coupling to torsion), the torsional Hall viscosity $\xi^\mathrm{H}$ vanishes: such a constraint would determine individual viscosities (and fix contact terms to be zero).  Looking further, we can use this knowledge to examine generic non-minimal responses to torsion, even in crystalline settings, where analogs of the $e \wedge T$ term have been studied\cite{else2021quantum}.  

\section{Modified Lamb surface waves: anisotropic viscosity} \label{sec:two}
\begin{figure}[t]
    \centering
    \includegraphics[width=.5\columnwidth]{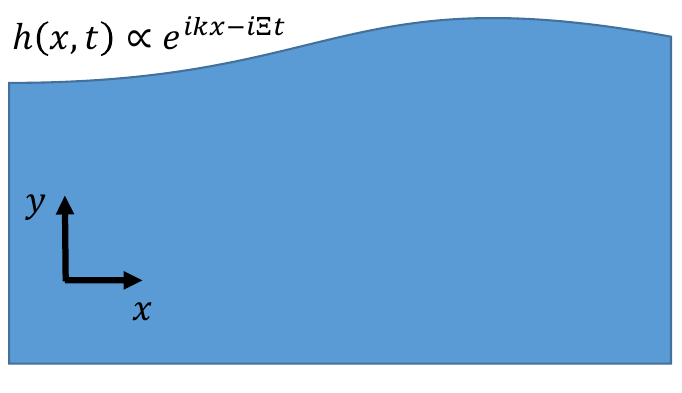}
    \caption{Half plane geometry where the height of the half plane is a surface wave with wavenumber $k$ and frequency $\Xi$. 
    Considering an anisotropic viscous fluid, we try to find the dispersion relation $\Xi(k)$.}
    \label{fig:halfplane}
\end{figure}

In this section we provide a more detailed derivation of the results of Sec.~\ref{sec:waves} for (incompressible) surface wave flow for a fluid with anisotropic odd viscosity in a half plane geometry, parameterized by $y = h(x,t)$ (see Figure.~\ref{fig:halfplane}). 
In particular, we would like to see how the dispersion $\Xi(k)$ of the surface waves is modified by the presence of our anisotropic odd viscosities, and how this is impacted by the dissipative and non-dissipative contact terms $C_0$ and $C_\mathrm{dis}$. 
We follow the strategy outlined in Ref.~\cite{abanov2018odd}, paying particular attention to the redundancies between the viscosity coefficients.  We choose to frame the velocity field in terms of potentials $\phi$ (velocity potential) and $\psi$ (stream function) such that $\psi$ is the only source of vorticity:
\beq
\label{eq:veltwopotssup}
v^i = \partial_i \phi + \epsilon^k_i \partial_k \psi.
\eeq
For the incompressible flow we consider, the velocity potential $\phi$ is harmonic
\beq
\label{eq:incompwave}
\boldsymbol{\nabla} \cdot {\bf v}= \nabla^2 \phi = 0.
\eeq
Similarly, the Laplacian of the stream function gives the vorticity
\beq
\boldsymbol{\nabla}\times\mathbf{v} = -\nabla^2\psi = \omega.
\eeq
In the bulk of the half plane, our viscous fluid must satisfy the momentum continuity equation--which serves as the bulk equation of motion:
\beq
D_t (\rho v_\mu) = \partial_t (\rho v_\mu) + \rho v_j \partial_j v_\mu = - \partial_j \tau_{\nu\mu}  -\rho     g\hat{y}_\mu.
\eeq
Here we have used the classical constitutive relation $\mathbf{g}_{\mathrm{mom}}=\rho\mathbf{v}$ to express the momentum density of the fluid in terms of the density $\rho$. 
We consider the Eulerian perspective of fluid flow and write the continuity equation in terms of a fluid derivative $D_t = \partial_t + v^i \partial_i$ \footnote{Alternatively, as in our previous work Ref.~\cite{rao2019hall}, we could have taken the Lagrangian perspective and group the non-linear convective term with the stress tensor in a momentum flux tensor\cite{lamb1924hydrodynamics,landau1986theory} }. 
As we are considering linearized surface waves for an incompressible fluid, we can set $\rho=1$ for convenience and neglect the higher-order convective term in the continuity equation to obtain the linearized equation of motion
\beq\label{eq:lineareom}
\partial_t {\bf v} = -\boldsymbol{\nabla} p + \eta^\mathrm{H}_{\mathrm{tot}} \boldsymbol{\nabla} \omega + \eta^{\mathrm{dis}}_{\mathrm{tot}}\;\boldsymbol{\nabla}^2 {\bf v} - g \hat{\bf y}.
\eeq
As expected, the  viscosities enter the equation of motion in terms of the sums $\eta^H_{\mathrm{tot}} = \eta^H +\bar{\eta}^H$ and $\eta^{\mathrm{dis}}_{\mathrm{tot}} = \eta^\mathrm{R} + \eta^\mathrm{sh}$. 
Here we notice that in the bulk non-dissipative viscosities can be thought of as a modification to the pressure of the fluid, in particular we can define the ``modified pressure'' \cite{abanov2018odd,ganeshan2017odd}
\beq
\tilde{p} = p - \eta^H_{\mathrm{tot}}\omega. 
\eeq
This is a manifestation of the ``triviality'' of the Hall viscosity in the bulk, since we can view it as a modification to the pressure of the fluid \cite{Wiegmann_boundary,Wiegmann_vortex}. 
We now see the equation of motion simplifies to 
\beq
\partial_t {\bf v} = -\boldsymbol{\nabla} \tilde{p} + \eta^{\mathrm{dis}}_{\mathrm{tot}}\;\boldsymbol{\nabla}^2 {\bf v} - g \hat{\bf y}.
\eeq
If we take the curl of the equation above, we find that the vorticity $\omega$ satisfies
\beq
\label{eq:vorticity}
\partial_t \omega = \eta^{\mathrm{dis}}_{\mathrm{tot}}\; {\boldsymbol{\nabla}}^2 \omega.
\eeq
The bulk equation of motion must be supplemented by boundary conditions, and for the problem at hand we are physically motivated\cite{lamb1924hydrodynamics} to choose a no-stress boundary condition at the surface of the half plane and a kinematic boundary condition on the velocity vector. 
These are, denoting the boundary as $Y = h(x,t)$:
\beq
\bal
\label{eq:bcs}
\hat{n}_\mu \tau_{\mu\nu}\biggr|_Y &= 0\\
v_y\bigg|_Y &=\partial_t Y. \;\;  
\eal
\eeq
These are sometimes referred to as the dynamic (stress condition) and kinematic (velocity condition) boundary conditions, respectively \cite{abanov2018odd}. 
We now have the equations of motion that are to be satisfied for our surface wave flow, and proceed by assuming a wave solution for the velocity potentials $\phi$ and $\psi$, where
\beq
\bal
\label{eq:waveansatz}
\phi = \left(-iA\frac{k}{|k|} e^{|k| y} + Be^{-|k| y}\right) e^{i k x - i\Xi t}\\
\psi = (C e^{my} + De^{-my}) e^{i k x - i\Xi t}.
\eal
\eeq
We enforce that the velocity be zero as $y \rightarrow -\infty$, meaning we need to set $B = 0$ and $D=0$ [$\mathrm{Re}(m)\geq 0$ by construction]. 
The incompressibility condition Eq.~\eqref{eq:incompwave} dictates that the wave-number $k$ parametrizes both the $x$ and $y$ dependence of the potential $\phi$, whereas $\psi$ requires two parameters $m$ and $k$. 
To begin to apply the boundary conditions in terms of the wave ansatz solutions in Eq.~\eqref{eq:waveansatz}, we explicitly write down the components of velocity according to Eq.~\eqref{eq:veltwopotssup},
\beq
\bal
\label{eq:velocityexpr}
v_x &= (A |k| e^{|k| y} + Cm e^{m y}) e^{i k x - i\Xi t},\\
v_y & = -i k (A e^{|k| y} + C e^{my}) e^{i k x - i\Xi t}.
\eal
\eeq
The physical velocity is determined by taking the real part of this expression. 
We see that the velocity potential $\phi$ appears through the coefficient $A$ and $\psi$ through $C$ -- consequently, the amplitude $C$ need be proportional to the vorticity. 
The kinematic boundary condition  tells us $\partial_t h = v_y(x,h,t)$ and thus the explicit behavior of the surface. 
This gives us the following relations for the height $h(x,t)$, the vorticity $\omega$ and the pressure $\tilde{p}$ from the velocity potentials
\beq
\bal
\label{eq:hvp}
h(x,t) &= \frac{k}{\Xi} (A + C) e^{i k x - i\Xi t},\\
\omega &= e^{i k x - i\Xi t} (k^2-m^2) C e^{my},\\
\tilde{p} &= \Xi \frac{k}{|k|} A e^{|k| y} e^{i k x - i\Xi t} - g y.
\eal
\eeq
The first expression comes from integrating the kinematic boundary condition with respect to time, and keeping only terms to lowest order in the wave amplitudes. 
The second expression comes from substituting our ansatz for the velocity into the definition of the vorticity. 
Finally, the third equation comes from writing Eq.~\eqref{eq:lineareom} in terms of $\phi$ and $\psi$, and making use of Eqs.~\eqref{eq:vorticity} and \eqref{eq:incompwave}.
We have thus reduced the problem to finding relations for $m$, $k$, $\Xi$ and the amplitudes $A$ and $C$, and move to apply the bulk equations of motion and no-stress boundary conditions. 
Our goal is find $\Xi$ as a function of $k$, and thus to find how the dispersion is affected by the viscosities and contact terms in our setup.
We first proceed by analyzing the bulk vorticity equation Eq.~\eqref{eq:vorticity}, 
\beq
\partial_t \omega = \eta^\mathrm{dis}_\mathrm{tot} \nabla^2 \omega.
\eeq
If we substitute our wave ansatz Eq.~\eqref{eq:hvp}, this leads to the relation
\beq
m^2 = k^2 - i\Xi/(\eta^\mathrm{sh} + \eta^R)
\eeq
between dispersion $\Xi$ and the parameters $m$ and $k$. 
The other two unknowns of the problem are the amplitude coefficients $A$ and $C$. 
The no stress boundary conditions in Eq.~\eqref{eq:bcs} should now supply us with enough information to estimate the dispersion and amplitudes for these surface waves. \\
In the bulk, we have the same setup as Lamb \cite{lamb1924hydrodynamics}, with the modified pressure $\tilde{p}$ playing the role of the pressure. 
On the boundary, the Hall viscosity has a contribution separate from the pressure and the resulting stress boundary conditions differ from Lamb's setup \cite{abanov2018odd,soni2019odd}. 
Further, our situation diverges further from previous works as our additional anisotropic Hall and dissipative viscosities ($\bar{\eta}^\mathrm{H}$ and $\eta^\mathrm{R}$) differentiate themselves from their usual counterparts ($\eta^\mathrm{H}$ and $\eta^\mathrm{sh}$) at the boundary. \\
We now unpack the no-stress conditions $f^{\text{bdry}}_j = \hat{n}_i \tau^i_{\hphantom{i}j} = 0$. 
In our linearized picture, the normal vector to the surface is $\hat{\mathbf{n}} \approx (0,1) = \hat{\bf y}$. 
Above linear order the normal vector depends on the function $h(x,t)$ and is non-constant. 
The statement that there is no stress at the boundary gives us two constraints-- first in the $y$ direction we have
\beq
\bal
\label{eq:dbc1}
f^{\text{bdry}}_y &= 0\\
\hookrightarrow p &= 2\eta^\mathrm{sh} \partial_y v_y - \eta^H (\partial_y v_x + \partial_x v_y) + \bar{\eta}^H \omega. \eal
\eeq
Using the explicit expressions for pressure, vorticity and velocity in Eqs.~\eqref{eq:velocityexpr} and \eqref{eq:hvp}, this condition becomes:
\beq
\bal
\label{eq:condition1}
A \left\lbrace \Xi^2  + 2\eta^{H} \Xi |k| k + 2 i \Xi k^2 \eta^\mathrm{sh}  - g |k| \right\rbrace + C \left \lbrace  2 i |k|\Xi \eta^\mathrm{sh} m + 2 k|k| \Xi \eta^H  
- g|k| \right\rbrace = 0.
\eal
\eeq
Surprisingly, the anisotropic  viscosities $\bar{\eta}^H$ and $\eta^R$ have cancelled out leaving a normal boundary condition identical to the cases considered in previous works \cite{abanov2018odd,soni2019odd}. 
Setting the tangential component of the boundary force to zero yields
\beq
\bal
\label{eq:dbc2}
f^{\text{bdry}}_x &= 0\\
\hookrightarrow 0 &= \eta^\mathrm{sh}(\partial_x v_y + \partial_y v_x) + \eta^H(\partial_y v_y - \partial_x v_x) - \eta^\mathrm{R} \omega.
\eal
\eeq
The anisotropic viscosities also do not enter this condition, which simplifies to:
\beq
\label{eq:cond2}
2A \left[\eta^\mathrm{sh} i k^2 +  \eta^H k|k|\right] + C\left[2\eta^H k m + 2 i k^2 \eta^\mathrm{sh} + \Xi \right]=0.
\eeq
We can combine the two conditions to form one overall {\it consistency condition} which relates $k$, the dispersion $\Xi$ and the viscosities. 
Since we are viewing $\Xi$ as a function of $\mathbf{k}$, and since the physical solutions are only determined by the real part of Eq.~\eqref{eq:velocityexpr}, we can restrict to $k>0$ without loss of generality; the $k<0$ solutions are obtained by complex conjugating our resultant expressions. 
Dividing Eq.~\eqref{eq:condition1} by \eqref{eq:cond2} gives, for $k>0$, 
\beq
\label{eq:constraint}
\frac{gk - \Xi^2 - 2\Xi k^2(\eta^H + i \eta^\mathrm{sh})}{2 k^2 (\eta^H + i \eta^\mathrm{sh})} = \frac{gk - 2 \Xi k (\eta^H k + i \eta^\mathrm{sh} m)}{\Xi + 2 k(\eta^H m+ i \eta^\mathrm{sh} k)}.
\eeq
We will use this equation to compute the dispersion $\Xi(k)$ in different limits, and examine how it is affected by the anisotropic viscosity and contact terms \footnote{We consider the case $k>0, \eta^H>0$ as the other cases are symmetric, see Ref.~\cite{ganeshan2017odd}},
Reorganizing Eq.~\eqref{eq:constraint}, and discarding a trivial solution with $m=k$ and $\Xi = 0$, we find a polynomial equation for $m$. 
Introducing dimensionless quantities, 
\beq
\bal
\beta^2 = \frac{(\eta^\mathrm{sh} + \eta^R) k^2}{\sqrt{gk}}, \; \alpha = \frac{\eta^{H}}{\eta^\mathrm{sh} + \eta^R}, \\ \kappa = \frac{m}{k} \;\; \text{and} \;\;  \gamma = \frac{\eta^\mathrm{sh}}{\eta^\mathrm{sh}+\eta^R}.
\eal
\eeq
We can now cast the consistency condition as
\beq 
\bal
\frac{\left[\kappa + 1 - 2 i\alpha \right]}{\beta^4} + (\kappa-1)^2(\kappa+1)^3 - 4 (\kappa^2 -1) (\alpha^2 + \gamma^2) +4 \gamma (\kappa-1)(\kappa+1)^2 - 2 i \alpha (\kappa - 1)(\kappa + 1 )^3= 0.
\eal
\eeq
\subsection{Gravity dominated waves}\label{sec:lambdispersiongamma}
We first consider the case where gravity dominates so $\beta << 1$, and rescale our coordinates to $x = \beta \kappa$ we find our constraint equation to be
\beq
\bal
x+\beta-2 i\alpha\beta + (x-\beta)^2(x + \beta)^3 - 4 \beta^3(x^2 - \beta^2)(\alpha^2 + \gamma^2) + 4\gamma(x - \beta)(x+\beta)^2 \beta^3 - 2 i\alpha\beta(x-\beta)(x+\beta)^3  = 0.
\eal
\eeq
\noindent \textbf{Zero viscosity solution.} The zero viscosity limit $\alpha = \beta = \gamma = 0$ gives the classical dispersion relation for gravity waves \cite{lamb1924hydrodynamics,ganeshan2017odd}
\beq
\Xi = \pm \sqrt{gk}.
\eeq
\noindent \textbf{Viscous corrections.} We now keep up to second order in $\beta$, representing small dissipative viscous corrections, and turn on a small non-dissipative correction $\alpha$. 
We keep the order of limits in analogy with Ref.~\cite{abanov2018odd}, the dissipative viscosities are smaller than $g$, i.e. that $\beta << 1$. 
The solution to the resulting constraint equation is given by \footnote{The linear term $B\beta$ in the dispersion vanishes ($B=0$)}
\beq
\bal
x_{\pm} &= A_{\pm} \beta^2 + C_{\pm} \\
C_{\pm} &=    e^{\mp i\pi/4}\\
A_+ &= \frac{e^{i\pi/4}}{2} \left[2\gamma - 2 i\alpha -1\right], \; \; A_- = \frac{e^{i\pi/4}}{2} \left[2\gamma - 2 i\alpha -i\right].
\eal
\eeq
The frequency in this case is given by:
\beq
\bal
\label{eq:disp}
\Xi_{\pm} &= \pm \sqrt{gk}-(2 i\gamma + \alpha)\eta^\mathrm{dis}_\mathrm{tot} k^2\\
&= \pm \sqrt{gk}- 2 i\eta^\mathrm{sh}k^2 - 2\eta^\mathrm{H}k^2.
\eal
\eeq
Despite the additional anisotropic viscosities in our picture, this result matches exactly the case where $\eta^R = \bar{\eta}^\mathrm{H} = 0$ considered in Ref.~\cite{ganeshan2017odd}. 
However we can now interpret this dispersion in terms of the total and differences between the viscosities:
\beq
\bal
\Xi_{\pm} &= \pm \sqrt{gk}- i\left(\eta^\mathrm{dis}_\mathrm{tot} + \eta^\mathrm{dis}_\mathrm{diff}\right) k^2 - \left(\eta^\mathrm{H}_\mathrm{tot} + \eta^\mathrm{H}_\mathrm{diff}\right) k^2.
\eal
\eeq
This dispersion is sensitive to both dissipative and non-dissipative contact terms, as the differences between odd viscosities and dissipative viscosities enter. 
To access the $k<0$ regime, we let $k\rightarrow |k|, \alpha \rightarrow - \alpha$ in Eq~\eqref{eq:disp} and find analogous solutions. 
\subsection{Pure (odd) viscosity waves: $g=0$}\label{sec:two_chiral}
We now consider the case where $g=0$ and the dynamics of our surface waves are dominated by viscosity. 
We also suppose that odd viscosity is playing the main role and $\eta^\mathrm{H} >> \eta^\mathrm{sh}, \eta^\mathrm{R}$ \footnote{we work with individual viscosities initially and then convert to bulk and boundary components}.  In this case, the constraint equation becomes
\beq
\frac{- \Xi^2 - 2\Xi k^2(\eta^H + i \eta^\mathrm{sh})}{2 k^2 (\eta^H + i \eta^\mathrm{sh})} = \frac{- 2 \Xi k (\eta^H k + i \eta^\mathrm{sh} m)}{\Xi + 2 k(\eta^H m+ i \eta^\mathrm{sh} k)}.
\eeq
This becomes (throwing out the trivial $\Xi = 0$ solution): 
\beq
\Xi^2 + 2\Xi k^2(\eta^\mathrm{H} + i \eta^\mathrm{sh}) + 2\Xi k (\eta^\mathrm{H} m + i \eta^\mathrm{sh} k) + 4 k^3 (\eta^\mathrm{H}+i\eta^\mathrm{sh})(\eta^\mathrm{H}-i\eta^\mathrm{sh})(m-k) = 0.
\eeq
If we utilize the relation $m^2 = k^2 - i\Xi/\eta^{\mathrm{dis}}_\mathrm{tot} \rightarrow \Xi = i(m-k)(m+k)\eta^\mathrm{dis}_\mathrm{tot}$, and throw out terms above first order in the dissipative viscosities we find
\beq
2 i \eta^\mathrm{dis}_\mathrm{tot} (m+k)^2 + 4 k^2 \eta^\mathrm{H} = 0.
\eeq
This leads to the following dispersion (keeping only the solution with $\mathrm{Re}(m)>0$ that decays into the bulk) 
\beq
\Xi = -2 \eta^\mathrm{H} k^2  - 2 i k^2 \sqrt{|\eta^\mathrm{H}| \eta^\mathrm{dis}_\mathrm{tot}}.
\eeq
The dispersion above describes chiral waves moving in a direction set by the odd viscosity. 
Importantly, it is only the $\textit{component}$ $\eta^\mathrm{H}$ rather than the full odd viscosity $\eta^H_{tot}$ that sets the direction. 
This means that the direction of these chiral waves cannot be determined from bulk data alone, or equivalently that the expression above is sensitive to the non-dissipative contact term \footnote{Because of the strong constraints on the size of the dissipative viscosities, we cannot make any clear judgement about the dependence on the dissipative contact term here}.

\section{Active fluids \& angular momentum conservation}\label{app:active}
In many classical chiral active fluids\cite{soni2019odd}, time reversal symmetry is broken by a local rotation rate $\Omega$ for fluid particles. 
In this case, for an isotropic and incompressible chiral active fluid, the stress takes on a modified form due to angular momentum conservation
\beq\label{eq:rotatingstress}
\tau^i_{\hphantom{i}j} = -p \delta^i_j + \eta^\mathrm{sh}\left(\partial^i v_j + \partial_j v^i \right) + \eta^\mathrm{H} (\partial^{*,i} v_j + \partial^i v^*_j) + \eta^\mathrm{R} \epsilon_{\mu\nu} (\omega - 2\Omega) + \bar{\eta}^\mathrm{H} \delta_{\mu\nu} (\omega - 2\Omega).
\eeq
We have effectively added two $\Omega$-dependent terms to our stress tensor. 
This corresponds to measuring vorticity of the fluid  in a locally rotating frame with frequency $\Omega$. 
We treat $\Omega$ as a fixed (constant) parameter of our setup, as in the physical situation of a colloidal chiral mixture \cite{soni2019odd}, and thus the modifications to the stress tensor do not enter the bulk equations of motion. 
On the boundary, however, the $\Omega$-dependent terms provide a steady-state boundary force
\beq
\label{eq:constbdryfrce}
f^\mathrm{bdry}_j = -2 \left(\eta^\mathrm{R} \hat{s}_j \Omega +\bar{\eta}^\mathrm{H} \hat{n}_j \Omega\right).
\eeq
The local rotation rate $\Omega$ causes an additional torque at the boundary due to $\eta^\mathrm{R}$ and an additional pressure contribution due to $\bar{\eta}^\mathrm{H}$. 
In what follows, we consider how this alternate form of time-reversal symmetry breaking could affect the viscous surface waves in Sec.~\ref{sec:two}. 
We also allow for a longitudinal friction from a substrate $f^\mathrm{fric}_j = -\mu v_j$ to be consistent with the experimental setup of Ref.~\cite{soni2019odd}. 
This term only enters the bulk equations of motion, and stabilizes a steady-state fluid velocity in the absence of external torques. 
We will analyze surface waves for this fluid both with and without gravity. 
To do so, we first begin by deriving the bulk equations of motion.
\subsection{Equations of motion}
The linearized continuity equation for momentum, again setting the density $\rho =1$ for convenience, is now given by
\beq\label{eq:omegabulkeoms}
\partial_t {\bf v} = - \nabla \tilde{p} + \eta^{\mathrm{dis}}_\mathrm{tot} \nabla^2 {\bf v} - g{\bf \hat{y}} - \mu {\bf v},
\eeq
where $\mu$ parametrizes the friction between the fluid and the substrate. 
Following the experimental considerations of Ref.~\cite{soni2019odd}, we have neglected the nonlinear term in the equations of motion.  Taking the curl of Eq.~\eqref{eq:omegabulkeoms} leads to the vorticity equation
\beq
\label{eq:vort2}                               
\partial_t \omega = \eta^\mathrm{dis}_\mathrm{tot} \nabla^2 \omega - \mu \omega.
\eeq
\subsection{Steady state flow}
The modifications we have made now allow for a steady-state vorticity (zeroth order in the amplitude of surface waves) whereas in previous setup in Sec. 
II with $\Omega = 0$ and $\mu = 0$ we necessarily had $\omega = 0$ at zeroth order. 
We can look to solve the vorticity equation in the steady state, where Eq.~\eqref{eq:vort2} becomes
\beq
(\eta^\mathrm{dis}_\mathrm{tot}\nabla^2 - \mu) \, \omega = 0.
\eeq
Again in the half plane geometry, $y\leq0$, it can be verified that
\beq
\label{eq:steadystate}
\omega_{s} = \frac{\eta^\mathrm{R}}{\eta^\mathrm{R} + \eta^\mathrm{sh}} (2\Omega) e^{y/\delta}
\eeq
satisfies the vorticity equation, where $\delta = ((\eta^\mathrm{R}+\eta^\mathrm{sh})/\mu)^{1/2}$ is the hydrodynamic length that appears in Ref.\cite{soni2019odd}. 
In choosing the multiplicative constant, we have anticipated the boundary conditions of Sec.~\ref{sec:omegabcs} below. 
The steady state vorticity corresponds to a flow profile in the $x$ direction (if there was a $y$ component, it would blow up as $x\rightarrow \infty$):
\beq
v_x = -\frac{\eta^\mathrm{R}}{\eta^\mathrm{R} + \eta^\mathrm{sh}} (2\Omega) \delta e^{y/\delta}.
\eeq
We refer to the zeroth order velocity at the boundary as $v_x^{(0)} \equiv v_x(y=0)$.
\subsection{Modification to surface wave boundary conditions}\label{sec:omegabcs}
We now consider the generalization of our earlier linearized surface wave boundary conditions to account for the presence of a steady state, zeroth order fluid velocity. 
In terms of our no-stress boundary condition we have, by expanding the normal vector and the stress tensor to first order,
\beq\label{eq:bcexpansion}
n_i \tau^i_{\hphantom{i}j} = (n^{(0)}_i+ \epsilon n^{(1)}_i) (\tau^{(0),i}_j + \epsilon \tau^{(1),i}_j) = 0.
\eeq
 Here $\epsilon\hat{\mathbf{n}}^{(1)}$ is the first order variation of the surface normal vector (taking into account the variations in the fluid height), and $\epsilon\tau_{\mu\nu}^{(1)}$ is the first order variation of the stress tensor (taking into account the linearized fluid velocity). 
 We consider our surface wave setup, where we treat the height $y = h(x,t)$ as a small perturbation around $y=0$. 
 This means that the normal vector can be written as
\beq\label{eq:normal}
{\bf \hat{n}} = {\bf \hat{n_0}} + \epsilon {\bf \hat{n_1}} \approx {\bf \hat{y}} - (\partial_x h) \bf{\hat{x}}.
\eeq
Collecting the zeroth order terms in Eq.~\eqref{eq:bcexpansion}, we have $n_\mu^{(0)} \tau^{(0)}_{\mu\nu} = 0$ and hence  
\beq
\bal
\label{eq:zeroth}
2\eta^\mathrm{R}\Omega - (\eta^\mathrm{sh} + \eta^\mathrm{R})\omega_s = 0\\
p_0 = \eta^\mathrm{H}_\mathrm{tot} \omega_s - 2\bar{\eta}^\mathrm{H} \Omega.
\eal
\eeq
The first equation is satisfied by our expression Eq.~\eqref{eq:steadystate} for the zeroth order vorticity. 
The second tells us that with the steady state vorticity Eq.~\eqref{eq:steadystate} we are able to set the steady state pressure outside of the half plane to $p_0 = \left(\frac{\eta^\mathrm{H} \eta^\mathrm{R}}{\eta^\mathrm{dis}_\mathrm{tot}} -\frac{\bar{\eta}^\mathrm{H} \eta^\mathrm{sh}}{\eta^\mathrm{dis}_\mathrm{tot}}\right) (2\Omega)$. 
At first order, we have  that $\epsilon n_\mu^{(1)}\tau^{(0)}_{\mu\nu} + \epsilon n_\mu^{(0)} \tau^{(1)}_{\mu\nu} = 0$. 
Inserting Eqs.~\eqref{eq:normal} and \eqref{eq:rotatingstress}, this gives
\beq
\bal
p_1 &= 2\eta^\mathrm{sh} \partial_y v_y - \eta^H (\partial_y v_x + \partial_x v_y) + \bar{\eta}^H \omega_1 + (\partial_x h)\left[\eta^\mathrm{dis}_\mathrm{diff} \omega_s + 2\Omega \eta^\mathrm{R} \right]  - h \partial_y(p_0-\eta^\mathrm{H}_\mathrm{tot} \omega_s)\\
0 &= \eta^\mathrm{sh}(\partial_x v_y + \partial_y v_x) + \eta^H(\partial_y v_y - \partial_x v_x) - \eta^\mathrm{R} \omega_1 + (\partial_x h) \left[\eta^\mathrm{H}_\mathrm{diff} \omega_s + p_0 + 2 \bar{\eta}^\mathrm{H} \Omega\right] - h \eta^\mathrm{dis}_\mathrm{tot} \partial_y \omega_s.
\eal
\eeq
We can apply the zeroth order boundary conditions to find
\beq
\bal
\label{eq:dispfriction}
p_1 &= 2\eta^\mathrm{sh} \partial_y v_y - \eta^H (\partial_y v_x + \partial_x v_y) + \bar{\eta}^H \omega_1 +2 (\partial_x h)\eta^\mathrm{sh} \omega_s \\ 
0 &= \eta^\mathrm{sh}(\partial_x v_y + \partial_y v_x) + \eta^H(\partial_y v_y - \partial_x v_x) - \eta^\mathrm{R} \omega_1 + 2 (\partial_x h)\eta^\mathrm{H} \omega_s - h \eta^\mathrm{dis}_\mathrm{tot} \partial_y \omega_s,
\eal
\eeq
where we have used the fact that from the zeroth-order boundary conditions $p_0-\eta^\mathrm{H}_\mathrm{tot}\omega_s$ is constant at the boundary.
The kinematic boundary condition in this case, where we have a zeroth order velocity, is given by
\beq
\label{eq:kbcfric}
\frac{d h}{d t}=\partial_t h + v_x^{(0)} \partial_x h = v_y(y=0,x,t).
\eeq
\subsection{Surface waves with $\Omega$}
We now continue on to consider surface waves with the time-reversal symmetry breaking coming from an internal rotation rate $\Omega$. 
The bulk vorticity equation is still
\beq
\partial_t \omega = \eta^\mathrm{dis}_\mathrm{tot} \nabla^2 \omega - \mu \omega.
\eeq
We can write the overall vorticity as a sum of the steady state contribution, which we just considered, and a contribution first-order in the amplitude of surface waves
\beq
\omega = \omega_s + \omega_1(x,y,t).
\eeq
To consider the first-order contribution to the vorticity, we again introduce velocity potentials that parameterize our surface wave Eq.~\eqref{eq:waveansatz}. 
The ansatz for the first order vorticity is then equivalent to Eq.~\eqref{eq:hvp} and is given by
\beq
\omega_1 = e^{i k x - i\Xi t} (k^2-m^2) C e^{my}.
\eeq
This satisfies the bulk equation of motion to linear order in the perturbative parameter
\beq
\partial_t \omega_1 = (\eta^\mathrm{dis}_\mathrm{tot} \nabla^2 - \mu) \omega_1.
\eeq
This leads to the modified condition
\beq
\bal
\label{eq:conditionfric}
\Xi = i\eta^\mathrm{dis}_\mathrm{tot}(m^2 - k^2) - i\mu.
\eal
\eeq
Our proposed form for the first order velocities and vorticities in Eq.~\eqref{eq:hvp} still hold. 
The bulk equation of motions mandate that the modified pressure now takes the form
\beq\label{eq:fricp}
\tilde{p}=p_1-\eta^\mathrm{H}_\mathrm{tot}\omega_1 - \mu \phi,
\eeq
which differs from Eq.~\eqref{eq:hvp} by the addition of $-\mu\phi$, where $\phi$ is the velocity potential. 
Eq.~\eqref{eq:waveansatz}. 
Additionally, the modified kinematic boundary condition Eq.~\eqref{eq:kbcfric} implies that the height $h(x,t)$ now takes the form
\beq
h(x,t) = \frac{v_y(y=0,x,t)}{-i\Xi(k) + i k v_x^{(0)}}\label{eq:frich}.
\eeq 
Now revisiting the first order boundary conditions Eq.~\eqref{eq:dispfriction}, we can substitute in our ansatz Eqs.~\eqref{eq:waveansatz}, \eqref{eq:fricp}, and \eqref{eq:frich} for the velocities, modified pressure, and height, respectively. 
The normal boundary condition in terms of surface wave parameters becomes
\beq
\bal
\label{eq:const1omega}
A\left[\Xi(k  v_x^{(0)}- \Xi) + gk + i\mu (k v_x^{(0)}-\Xi) + 2 k^2 (k v_x^{(0)} - \Xi) + 2 k^2 (k v_x^{(0)} - \Xi)(\eta^\mathrm{H}+i\eta^\mathrm{sh} + 2 i\eta^\mathrm{sh}) \omega_s k^2 \right] \\
+ C \left[gk + 2 k(\eta^\mathrm{H} k + i\eta^\mathrm{sh} m) (k v_x^{(0)}- \Xi) + 2 i \eta^\mathrm{sh} \omega_s k^2 \right] = 0.
\eal
\eeq
The tangential boundary condition becomes
\beq
\bal
\label{eq:const2omega}
A\left[2(kv_x^{(0)} - \Xi)k^2(\eta^\mathrm{H} + i\eta^\mathrm{sh}) + 2 k^2\eta^\mathrm{H} \omega_s + \eta^\mathrm{dis}_\mathrm{tot} k \partial_y \omega_s \right] \\+ C\left[(\Xi - i\mu)(kv_x^{(0)}-\Xi) + 2(kv_x^{(0)}-\Xi)k(\eta^\mathrm{H} m + i \eta^\mathrm{sh} k) + 2 k^2\eta^\mathrm{H} \omega_s + \eta^\mathrm{dis}_\mathrm{tot} k\partial_y \omega_s\right]=0.
\eal
\eeq
The equations above Eq.~\eqref{eq:const1omega} and Eq.~\eqref{eq:const2omega} represent our consistency conditions for the wave setup with $\Omega$ and $\mu$. 
To solve the consistency conditions, we can combine Eqs.~\eqref{eq:const1omega} and \eqref{eq:const2omega} with Eq.~\eqref{eq:conditionfric} to find three nontrivial solutions for $m(k)$ that can have $Re(m)>0$. 
Due to the complicated nature of the consistency condition, to make progress we will focus analytically on three cases. 
First, we will consider surface waves in the limit of long-wavelength  $k\delta\ll 1$ and zero gravity. 
Second, we will keep $k\delta\ll 1$ and introduce gravity as a small perturbation $g\delta\ll\eta_\mathrm{tot}^\mathrm{dis}\Omega$. 
Third, we will consider the large gravity limit.

\subsubsection{$g=0$}
We first consider the case without gravity, which was the setup in Ref.~\cite{soni2019odd}. 
In this case, in the long wavelength $k\delta << 1$ limit, there are two modes which always decay into the bulk. 
The first is, to third order, 
\beq
\bal
\Xi_{1,g=0} =2(i\eta^\mathrm{H} - \eta^\mathrm{sh}) \frac{2\Omega \delta \eta^\mathrm{R}}{\mu \eta^\mathrm{dis}_\mathrm{tot}} k^3 + \mathcal{O}[(k\delta)^{5/2}].
\eal
\eeq
This mode matches exactly that found in the corresponding long wavelength limit in Ref.\cite{soni2019odd}, despite the addition of the additional Hall viscosity $\bar{\eta}^\mathrm{H}$ \footnote{Our rotational viscosity has an opposite sign by definition and we define viscous force as the divergence on the first index of the stress tensor rather than the second, leading to additional sign differences, however our results are substantively compatible}. 
It leads directly to the stability condition $\text{sign}(\eta^\mathrm{H} \eta^\mathrm{R} \Omega)<0$ in order for perturbations to decay in time. 
Additionally, there is always an overdamped excitation with dispersion given by
\beq
\Xi_{2,g=0} = -i\mu - \frac{2\eta^\mathrm{R} \Omega}{\eta^\mathrm{dis}_\mathrm{tot}} k\delta + e^{i\pi/4} (\eta^\mathrm{H} + i\eta^\mathrm{sh}) \sqrt{\frac{2\Omega \eta^\mathrm{R}}{\mu (\eta^\mathrm{dis}_\mathrm{tot})^{3/2}}} k^{3/2} +\mathcal{O}[(k\delta)^2].
\eeq
This solution is effectively dominated by damping due to the friction term in the limit $k\delta << 1$. 
We will see below, however, that for nonzero $g$ this mode is essential to recovering the second branch of our Lamb wave solutions Eq.~\eqref{eq:disp}.
Finally, there is a third nontrivial solution that can decay into the bulk. 
It corresponds to the solution
\beq
m_{3,g=0}(k)=\frac{k\eta_\mathrm{diff}^\mathrm{dis}}{\eta_\mathrm{tot}^\mathrm{dis}},
\eeq
which decays into the bulk whenever $\eta^\mathrm{R}\le \eta^\mathrm{sh}$. 
The dispersion relation is
\beq
\Xi_{3,g=0}(k)=-i\mu-4 i\frac{\eta^\mathrm{R}\eta^\mathrm{sh}k^2}{\eta_\mathrm{tot}^\mathrm{dis}}+\mathcal{O}[(k\delta)^3].
\eeq
This mode is overdamped and almost completely stationary at small $k\delta$. 
We will see below that this mode is always unphysical for $g$ large enough (or equivalently, for $\mu$ small enough).
\subsubsection{Small gravity case}
We now consider the case where gravity is small and again with the long wavelength limit $k\delta << 1$. 
For the two main physical modes, we find that the effect of gravity is, to lowest order, to introduce a linear in $k$ correction to the damping rate, given by 
\beq
\bal
\Xi_{1g}(k) &= \Xi_{1,g=0}(k) - \frac{i g k \delta}{\sqrt{\eta^\mathrm{dis}_\mathrm{tot} \mu}} + ... \\
\Xi_{2 g}(k) &=\Xi_{2,g=0}(k)  + \frac{i g k \delta}{\sqrt{\eta^\mathrm{dis}_\mathrm{tot} \mu}} + ...
\eal
\eeq
The effect of gravity is more drastic on the $\Xi_3$ mode. 
First, we find that to linear order in $g$, $m_3(k)$ is given by
\begin{equation}
    m_3(g) = \frac{k}{\eta^\mathrm{dis}_\mathrm{tot}}\left({\eta_\mathrm{diff}^\mathrm{dis}} + \frac{\eta^\mathrm{H}\delta g}{\eta^\mathrm{R}\Omega}\right).
\end{equation}
Stability of the fluid requires the second term to be strictly negative. 
This implies that the $\Xi_3$ mode will become unphysical even for small $g$, provided $\eta^\mathrm{H}$ and $1/\eta^\mathrm{R}$ are large enough. 
As such, we will neglect the $\Xi_3$ mode in what follows.
\subsubsection{Gravity $g\neq 0$ case}\label{sec:gneq0}
\begin{figure}[ht]
\subfloat[]{
\includegraphics[width=0.4\textwidth]{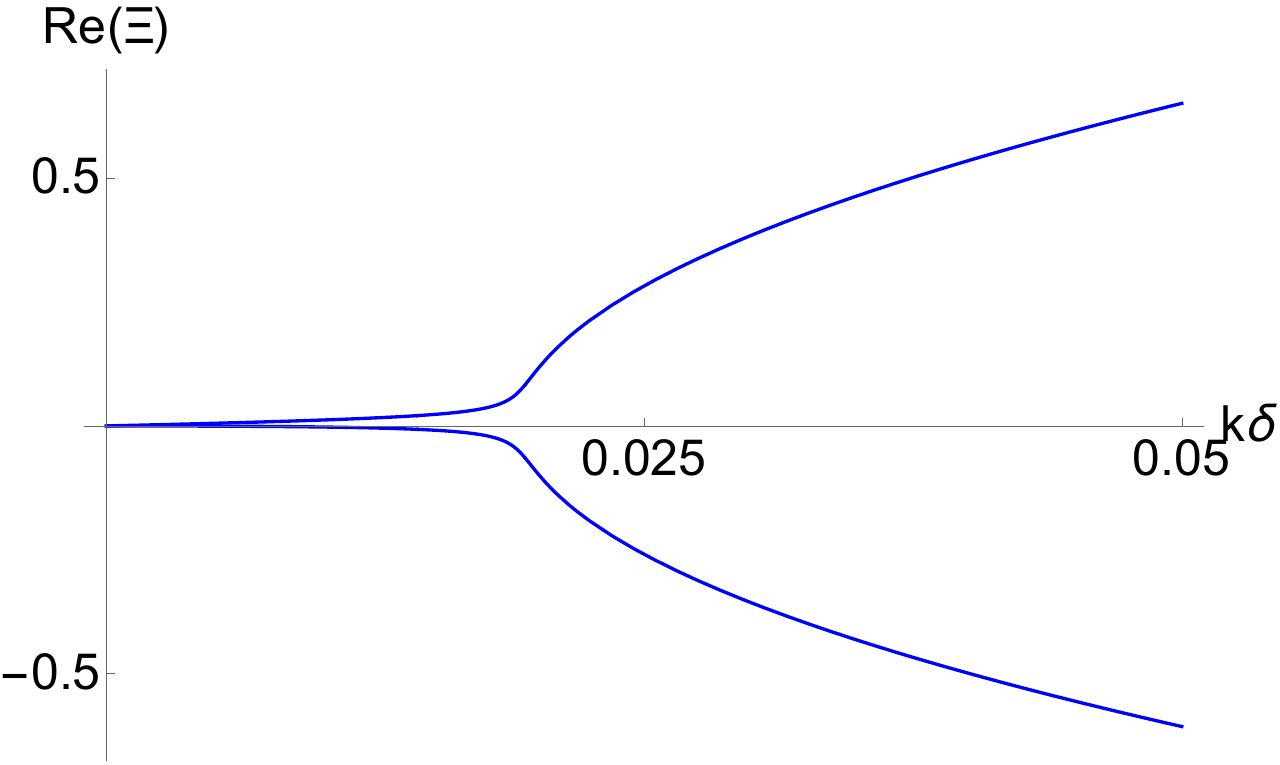}
}
\quad
\subfloat[]{
\includegraphics[width=0.4\textwidth]{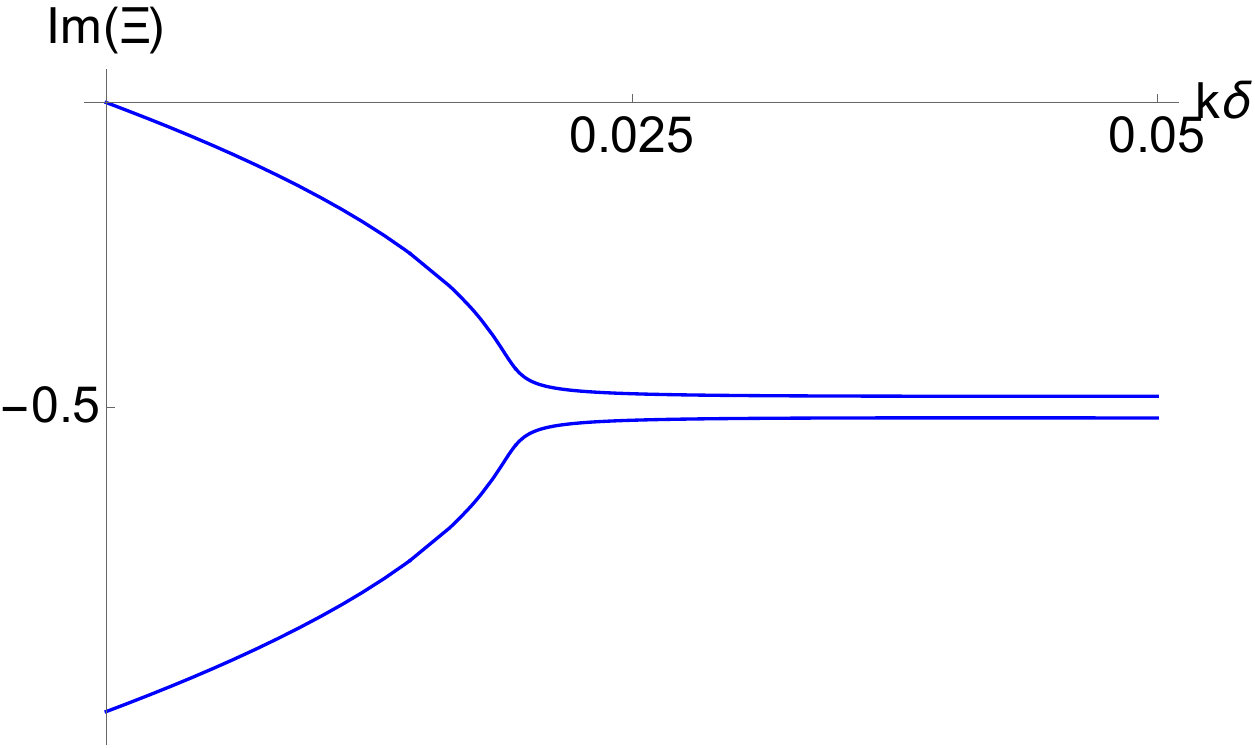}
}
\caption{Dispersion (a) and Damping (b) for the modes $\Xi_{g1}$ and $\Xi_{g2}$ with $\eta^\mathrm{sh}=0.1,\eta^\mathrm{R}=0.5,\eta^\mathrm{H}=0.3, \mu=1,\omega_s=-1$ and $g=10$. 
There is a crossover from friction-dominated behavior at $k\delta\lesssim 0.025$ to Lamb wave-like behavior at $k\delta \gtrsim 0.025$.}\label{fig:dispersion}
\end{figure}

To examine the surface waves for general $g$ and $k$, let us first return to the consistency conditions Eqs.~\eqref{eq:const1omega} and \eqref{eq:const2omega}. 
Note that for $\omega_s,\mu\rightarrow 0$, this reproduces exactly the consistency equation we obtained for gravity-dominated Lamb waves in Eq.~\eqref{eq:constraint}. 
We thus expect that when $g\delta \gg \eta^\mathrm{dis}_\mathrm{tot}\Omega$, we should recover the two branches of the modified Lamb wave dispersion Eq.~\eqref{eq:disp}. 
We examine the two modes $\Xi_{1g}(k)$ and $\Xi_{2g}(k)$ in the limit of large $g\delta/\eta_\mathrm{dis}^\mathrm{tot}\Omega$. 
We expect that $\Xi_{1g} \sim -\sqrt{gk}$ and $\Xi_{2g} \sim \sqrt{gk}$ as $\Omega \rightarrow 0$. 
To see how this occurs, we show in Fig.~\ref{fig:dispersion} the real and imaginary parts of $\Xi_{1,2}$ for generic values $\eta^\mathrm{sh}=0.1,\eta^\mathrm{R}=0.5,\eta^\mathrm{H}=0.3, \mu=1,\omega_s=-1$ with $g=10$. 
We see in Fig.~\ref{fig:dispersion}(a) that for $\mathrm{Re}(\Xi)$ there is a crossover from nearly stationary behavior at small $k$ to a dispersion consistent with $\mathrm{Re}(\Xi)\sim\pm\sqrt{gk}$ at larger $k$. 
In Fig.~\ref{fig:dispersion}(b) we see that the damping rate $\mathrm{Im}(\Xi)$ for the two modes depend linearly on $k$ for small $k$, and are approximately equal at larger $k$, varying as $\mathcal{O}(k^2)$. 
Expanding $\Xi_{1g}$ and $\Xi_{2g}$ to lowest order in $k\delta$ captures the behavior of the dissipation at small $k$, yielding
\beq
\bal
\Xi_{1g}(k) &= -\frac{i g k}{\mu} + ... \\
\Xi_{2 g}(k) &= -i\mu + \frac{i g k}{\mu} - \frac{2\eta^\mathrm{R} \Omega}{\eta^\mathrm{dis}_\mathrm{tot}} k\delta + ...
\eal
\eeq
Next, we can analyze the dispersion asymptotically for large $g$. 
First, note that when both the dissipative and Hall viscosities are zero, the flow is pure potential flow (as in the case $\Omega=0$). 
In this limit, we find the viscosity-free dispersion relation
\begin{equation}
    \Xi_{0} = -\frac{i\mu}{2}\pm\frac{1}{2}\sqrt{4gk-\mu^2},\label{eq:xi0friction}.
\end{equation}
This describes propagating damped waves for $k$ greater than the threshold wavevector $k_*=\mu^2/(4g)$, and overdamped stationary waves for $k<k_*$. 
In analogy with Sec.~\ref{sec:lambdispersiongamma}, we can compute the dispersion perturbatively for small $\beta=\sqrt{\eta_\mathrm{tot}^\mathrm{dis}k^2}/(gk)^{1/4}$, which corresponds to a large-$g$ expansion. 
In full analogy with our modified Lamb waves of Sec.~\ref{sec:lambdispersiongamma}, we find
\begin{equation}
    \Xi_{g\rightarrow\infty} = \pm\sqrt{gk} -\frac{i\mu}{2} - 2k^2(\eta^\mathrm{H}+i\eta^\mathrm{sh}) -\frac{1}{2}k\delta\omega_s.
\end{equation}
The first two terms correspond to the first two terms in the Taylor expansion of $\Xi_0$ in Eq.~\eqref{eq:xi0friction} for large $g$. 
The second term is identical to the modification to the Lamb wave dispersion found in Sec.~\ref{sec:lambdispersiongamma}. 
Finally, the last term gives the correction to the dispersion due to the nonzero angular velocity $\Omega$ of the fluid particles. 
This matches with our observations in Fig.~\ref{fig:dispersion}.
\begin{figure}
    \centering
    \includegraphics[scale=.5]{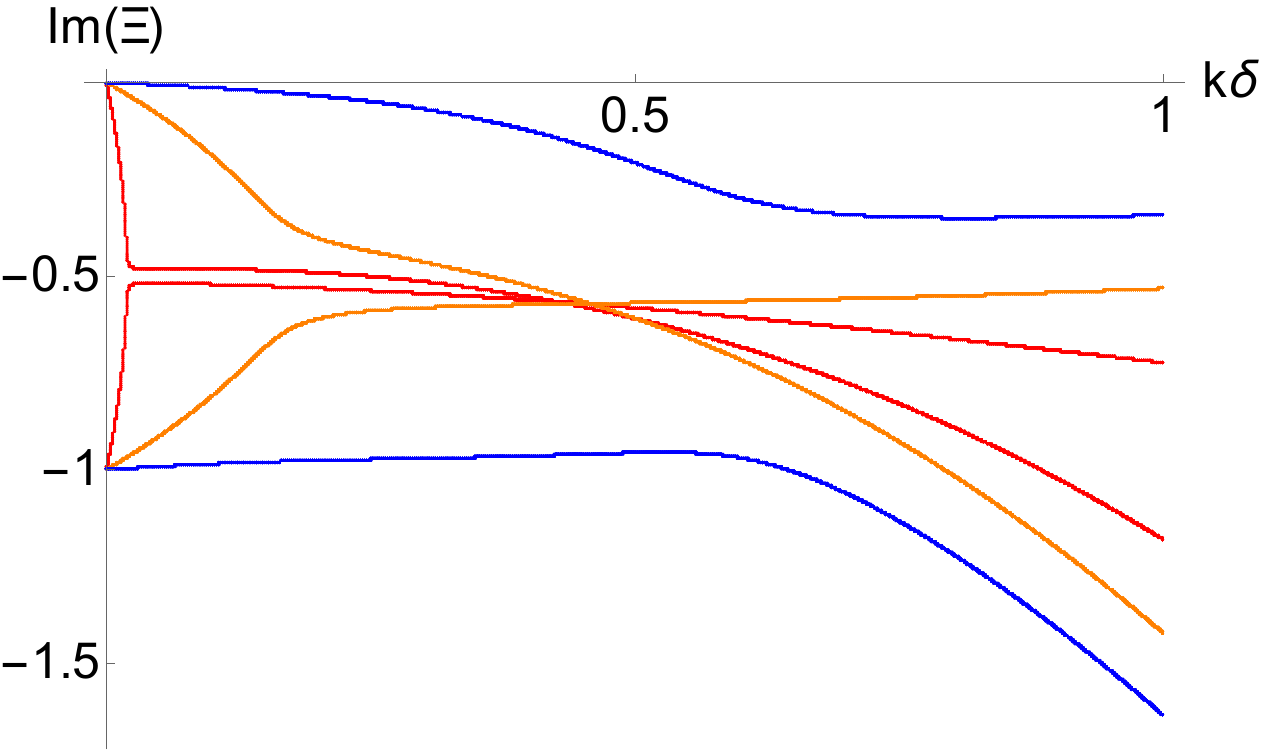}
    \caption{Corresponding damping $\Im(\Xi)(k)$ for surface waves with gravity with time reversal breaking from a local rotation rate $\Omega$ to accompany Fig,~\ref{fig:dispgneq0}. 
    The red plot has $g=10$, the blue plot has $g=1$ and the orange has $g=1.2$. 
    The other parameters are fixed at $\eta^\mathrm{sh}=0.1,\eta^\mathrm{R}=0.5,\eta^\mathrm{H}=0.3$ and $\mu=1$.}
    \label{fig:dissipation}
\end{figure}
Lastly, in Fig.~\ref{fig:dissipation} we show the imaginary part of $\Xi_{1,2g}$ for the three different values of $g$ discussed in Sec.~\ref{sec:wave_fluid}. 
We see that for small $k$, the damping rate for $\Xi_{1g}$ always goes to zero, while the damping rate for $\Xi_{2g}$ always goes to $\mu$.
\twocolumngrid
\bibliography{boundary_fluids}
\end{document}